\documentclass[sigconf]{acmart}

\usepackage{amsmath,amsfonts}
\usepackage{algorithm}
\usepackage{graphicx}
\usepackage{textcomp}
\usepackage{makecell}
\usepackage{colortbl}
\usepackage{adjustbox}
\usepackage{color}
\usepackage{xcolor}
\usepackage[caption = false]{subfig}
\usepackage{hyperref}
\usepackage{amsthm}
\usepackage{microtype}      
\usepackage{algpseudocode}  
\usepackage{soul}
\usepackage{hyperref}
\usepackage{url}
\usepackage{bbm}

\setcopyright{none} 
\acmConference[Anonymous Submission to ACM CCS 2019]{ACM Conference on Computer and Communications Security}{Due 15 May 2019}{London, TBD}
\acmYear{2019}

\begin{document}
\title{Back in Black: A Comparative Evaluation of Recent State-Of-The-Art Black-Box Attacks} 

\author{Kaleel Mahmood}
\email{kaleel.mahmood@uconn.edu}
\affiliation{%
 \streetaddress{Department of Computer Science and Engineering}
 \institution{University of Connecticut, USA}
 }

\author{Rigel Mahmood}
\affiliation{%
 \streetaddress{Department of Computer Science and Engineering}
 \institution{University of Connecticut, USA}
 }

\author{Ethan Rathbun}
\affiliation{%
 \streetaddress{Department of Computer Science and Engineering}
 \institution{University of Connecticut, USA}
 }

\author{Marten van Dijk}
\affiliation{%
 \institution{CWI, The Netherlands}
 }

\begin{abstract}
The field of adversarial machine learning has experienced a near exponential growth in the amount of papers being produced since 2018. This massive information output has yet to be properly processed and categorized. In this paper, we seek to help alleviate this problem by systematizing the recent advances in adversarial machine learning black-box attacks since 2019. Our survey summarizes and categorizes 20 recent black-box attacks. We also present a new analysis for understanding the attack success rate with respect to the adversarial model used in each paper. Overall, our paper surveys a wide body of literature to highlight recent attack developments and organizes them into four attack categories: score based attacks, decision based attacks, transfer attacks and non-traditional attacks. Further, we provide a new mathematical framework to show exactly how attack results can fairly be compared.
\end{abstract}

\ccsdesc{Security and privacy;~Computing methodologies;Machine Learning}

\keywords{Adversarial machine learning; adversarial attack; black-box attack}

\maketitle
\section{Introduction}

\begin{figure*}
\centering
\includegraphics[scale=0.55]{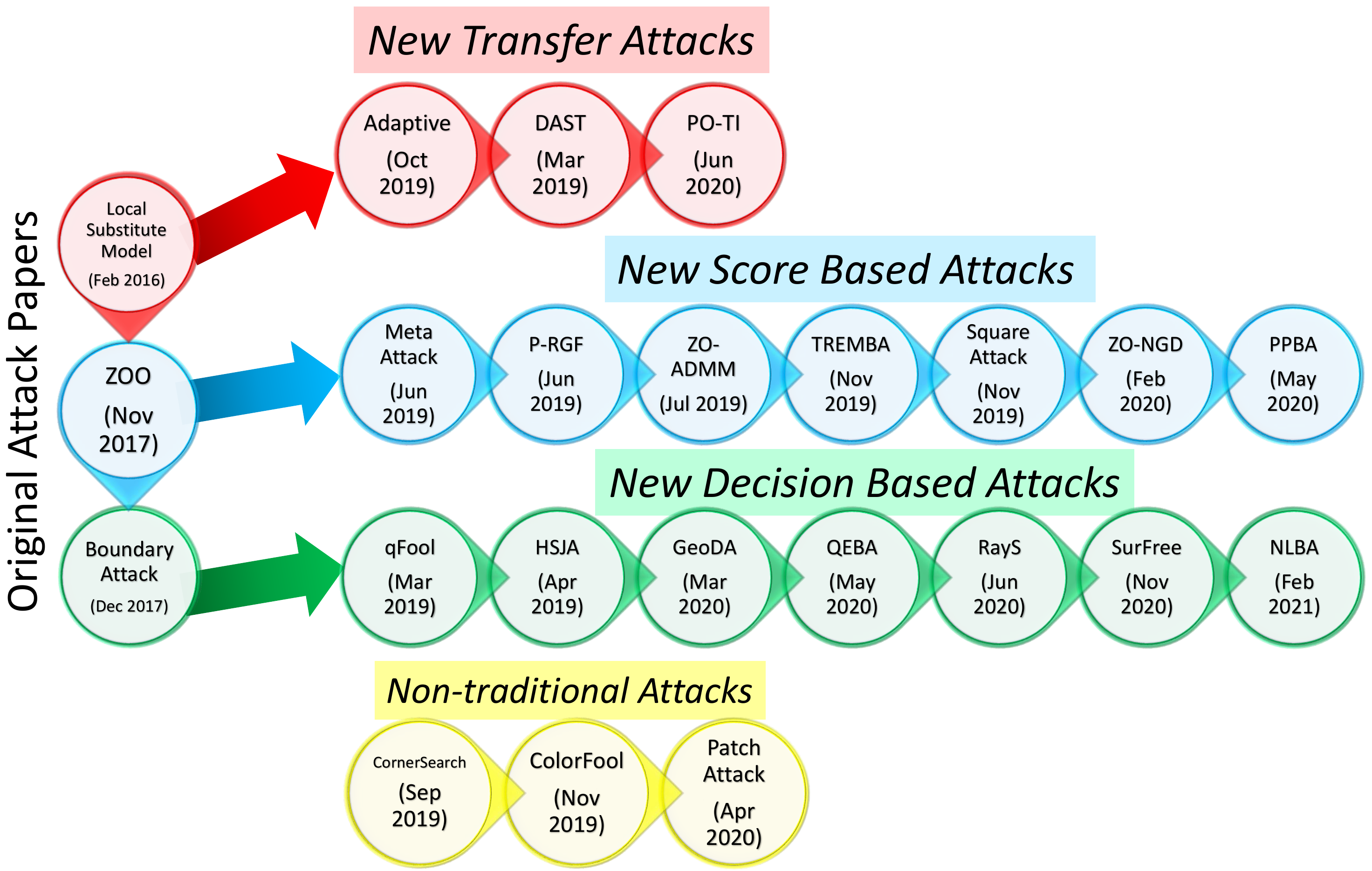}
\caption[]{Timeline of recent black-box attack developments. The transfer based attacks are show in red. The original transfer attack (Local Substitute Model) was proposed in~\cite{papernot2017practical}. The score based attacks are shown in blue. One of first widely adopted score based attacks (ZOO) was proposed in~\cite{chen2017zoo}. The decision based attacks are shown in green. One of the first decision based attacks (Boundary Attack) was proposed in~\cite{brendel2017decision}.}
\vspace{2mm}
\noindent
\label{fig:timeline}
\end{figure*}

One of the first works to popularize Convolutional Neural Networks (CNN)~\cite{FirstCNN} for image recognition was published in 1998. Since then, CNNs have been widely employed for tasks like image segmentation~\cite{He2017}, object detection~\cite{Redmon2016} and image classification~\cite{Kolesnikov2020}. Although CNNs are the de facto choice for machine learning tasks in the imaging domain, they have been shown to be vulnerable to adversarial examples~\cite{goodfellow2014explaining}. In this paper, we discuss adversarial examples in the context of images. Specifically, an adversarial example is an input image which is visually correctly recognized by humans, but has a small noise added such that the classifier (i.e. a CNN) misclassifies the image with high confidence. 

Attacks that create adversarial examples can be divided into two basic types, white-box and black-box attacks. White-box attacks require knowing the structure of the classifier as well as the associated trained model parameters~\cite{goodfellow2014explaining}. In contrast to this, black-box attacks do not require directly knowing the model and trained parameters. Black-box attacks rely on alternative information like query access to the classifier~\cite{chen2017zoo}, knowing the training dataset~\cite{ papernot2017practical}, or transferring adversarial examples from one trained classifier to another~\cite{dast}.

\begin{figure*}
\centering
\includegraphics[scale=0.75]{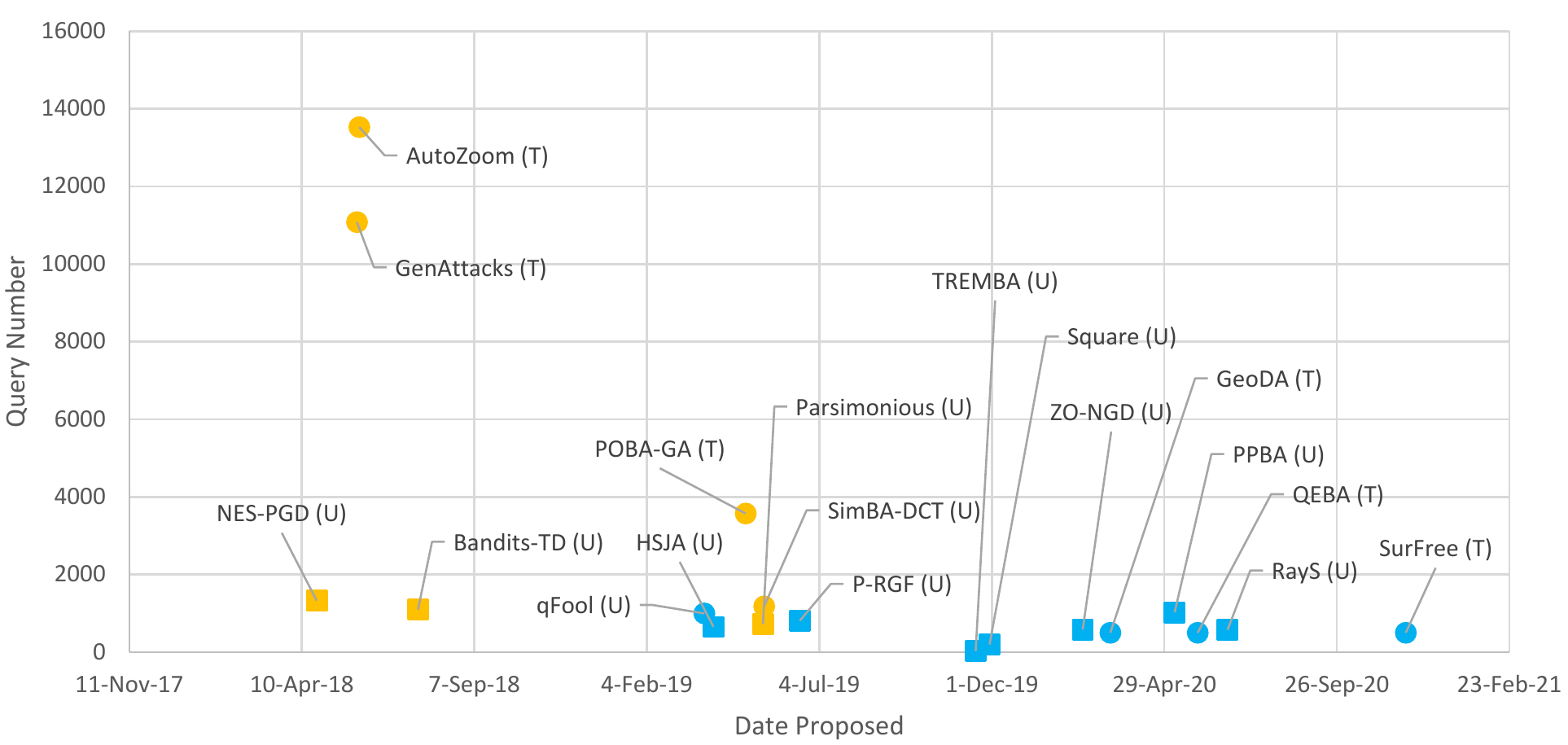}
\caption[]{Graph of different black-box attacks with the respective date they were proposed (e-print made available). The query number refers to the number of queries used in the attack on an ImageNet classifier. The orange points are attacks covered in previous survey work~\cite{bhambri2019survey}. The blue points are attacks covered in \textit{this} work. We further denote whether the attack is targeted or untargeted by putting a U or T next to the text label in the graph. A square point represents an attack done with respect to the $l_{2}$ norm and a circular point represents attacks done with respect to the $l_{\infty}$ norm.}
\vspace{2mm}
\noindent
\label{fig:OldVsNew}
\end{figure*}

In this paper, we survey recent advances in black-box adversarial machine learning attacks. We select this scope for two main reasons. First, we choose the black-box adversary because it represents a realistic threat model where the classifier under attack is not directly visible. It has been noted that a black-box attacker represents a more practical adversary~\cite{chen2020hopskipjumpattack} and one which corresponds to real world scenarios~\cite{papernot2017practical}. The second reason we focus on black-box attacks is due to the large body of recently published literature. As shown in Figure~\ref{fig:timeline}, many new black-box attack papers have been proposed in recent years. These attacks are not included in current surveys or systematization of knowledge papers. Hence, there is a need to categorize and survey these works, which is precisely the goal of this paper. To the best of our knowledge, the last major survey~\cite{bhambri2019survey} on adversarial black-box attacks was done in 2020. A graphical overview of the coverage of some of the new attacks we provide (versus the old attacks previously covered) are shown in Figure~\ref{fig:OldVsNew}. The complete list of important attack papers we survey are graphically shown in Figure~\ref{fig:timeline} and also listed in Table~\ref{tbl:mainTable}.

While each new attack paper published contributes to the literature, they often do not compare with other state-of-art techniques, or adequately explain how they fit within the scope of the field. In this survey, we summarize 20 recent black-box attacks, categorize them into four basic groups and create a mathematical framework under which results from different papers can be compared.    
\subsection{Advances in Adversarial Machine Learning}
In this subsection we briefly discuss the history and development of the field of adversarial machine learning. Such a perspective helps illuminate how the field went from a white-box attack like FGSM~\cite{goodfellow2014explaining} in 2014 which required complete knowledge of the classifier and trained parameters, to a black-box attack in 2021 like SurFree~\cite{surfree} which can create an adversarial example with only query access to the classifier using 500 queries or less. 

The inception point of adversarial machine learning can be traced back to several source papers. However, identifying the very first adversarial machine learning paper is a difficult task as the first paper in the field depends on how the term "adversarial machine learning" itself is defined. If one defines adversarial machine learning as exclusive to CNNs, then in~\cite{szegedy2013intriguing} the vulnerability of CNNs to adversarial examples was first demonstrated in 2013. However, others~\cite{Biggio2018} claim adversarial machine learning can be traced back as early as 2004. In~\cite{Biggio2018}, the authors claim evading linear classifiers which constituted email spam detectors was one of the first examples of adversarial machine learning. 

Regardless of the ambiguous starting point of adversarial examples, it remains a serious open problem which occurs across multiple machine learning domains including image recognition~\cite{goodfellow2014explaining} and natural language processing~\cite{hsieh2019robustness}. Adversarial machine learning is also not just limited to neural networks. Adversarial examples have been shown to be problematic for decision trees, k-nearest neighbor classifiers and support vector machines~\cite{papernot2016transferability}.

The field of adversarial machine learning with respect to computer visions and imaging related tasks, first developed with respect to white-box adversaries. One of the first and most fundamental attacks proposed was the Fast Gradient Sign Method (FGSM)~\cite{goodfellow2014explaining}. In the FGSM attack, the adversary uses the neural network model architecture $F$, loss function $L$, trained weights of the classifier $w$ and performs a single forward and backward pass (backpropagation) on the network to obtain an adversarial example from a clean example $x$. Subsequent work included methods like the Projected Gradient Descent (PGD)~\cite{madry2018towards} attack, which used multiple forward and backward passes to better fine tune the adversarial noise. Other attacks were developed to better determine the adversarial noise by forming an optimization problem with respect to certain $l_{p}$ norms, such as in the Carlini $\&$ Wagner~\cite{carlini2017towards} attack, or the Elastic Net attack~\cite{chen2018ead}. Even more recent attacks~\cite{croce2020reliable} have focused on breaking adversarial defenses and overcoming false claims of security which are caused by a phenomena known as gradient masking~\cite{athalye18a}.   

All of the aforementioned attacks are considered white-box attacks. That is, the adversary requires knowledge of the network architecture $F$ and trained weights $w$ in order to conduct the attack. Creating a less capable adversary (i.e., one that did not know the trained model parameters) was a motivating factor in developing black-box attacks. In the next subsection, we discuss black-box attacks and the categorization system we develop in this paper. 

\subsection{Black-box Attack Categorization}
We can divide black-box attacks according to the general adversarial model that is assumed for the attack. The four categories we use are transfer attacks, score based attacks, decision based attacks and non-traditional attacks. We next describe what defines the different categorizations and also mention the primary original attack paper in each category.  

\textbf{Transfer Attacks:} One of the first of black-box attacks was called the local substitute model attack~\cite{papernot2017practical}. In this attack, the adversary was allowed access to part of the original training data used to train the classifier, as well as query access to the classifier. The idea behind this attack was that the adversary would query the classifier to label the training data. After this was accomplished, the attacker would train their own independent classifier, which it is often referred to as the \textit{synthetic model}~\cite{mahmood2019buzz}. Once the synthetic model was trained, the adversary could run any number of white-box attacks on the synthetic model to create adversarial examples. These examples were then submitted to the unseen classifier in the hopes the adversarial examples would \textit{transfer} over. Here transferability is defined in the sense that adversarial examples that are misclassified by the synthetic model will also be misclassified by the unseen classifier. 

Recent advances in transfer based attacks include not needing the original training data like in the DaST attack~\cite{dast} and using methods that generate adversarial example with higher transferability (Adaptive~\cite{mahmood2019buzz} and PO-TI~\cite{towards_transfer}).

\textbf{Score Based Attacks:} The zeroth order optimization based black-box attack (ZOO)~\cite{chen2017zoo} was one of the first accepted works to rely on a query based approach to creating adversarial examples. Unlike transfer attacks which require a synthetic model, score based attacks repeatedly query the unseen classifier to try and craft the appropriate adversarial noise. As the name implies, for score based attacks to work, they require the output from the classifier to be the score vector (either probabilities or in some cases the pre-softmax logits output). 

Score based attacks represent an improvement over transfer attacks in the sense that no knowledge of the dataset is needed since no synthetic model training is required. In very broad terms, the recent developments in score based attacks mainly focus on reducing the number of queries required to conduct the attack and/or reducing the magnitude of the noise required to generate a successful adversarial example. New score based attacks include qMeta~\cite{Du2020Query-efficient}, P-RGF~\cite{cheng2019improving},  ZO-ADMM~\cite{zhao2019design}, TREMBA~\cite{huang2019black}, Square attack~\cite{andriushchenko2020square}, ZO-NGD~\cite{zhao2020towards} and PPBA~\cite{li2020projection}.   

\textbf{Decision Based Attacks:} We consider the type of attack that does not rely on a synthetic model and does not require the score vector output to be a decision based attack. Compared to either transfer based or score based attacks, decision based attacks represent an even more restricted adversarial model, as only the hard label output from the unseen classifier is required. The first prominent decision based attack paper was the Boundary Attack~\cite{brendel2017decision}. Since then, numerous decision based attacks have been proposed to improve upon the number of queries to successfully attack the unseen classifier, or reduce the noise required in the adversarial examples. The new decision attacks we cover in this paper include qFool~\cite{qfool}, HSJA~\cite{chen2020hopskipjumpattack}, GeoDA~\cite{rahmati2020geoda}, QEBA~\cite{qeba}, RayS~\cite{chen2020rays}, SurFree~\cite{surfree} and NonLinear-BA~\cite{nonlinear_ba}.

\textbf{Non-traditional Attacks:} The last category of attacks that we cover in this paper are called non-traditional black-box attacks. Here, we use this category to group the attacks that do not use standard black-box adversarial models. Transfer based attacks, score based attacks, and decision based attacks typically focus on designing the attack with respect the $l_{2}$ and/or the $l_{\infty}$ norm. Specifically, these attacks either directly or indirectly seek to satisfy the following condition: $||x-x_{adv}||_{p} \leq \epsilon$ where $x$ is the original clean example, $\epsilon$ is the maximum allowed perturbation and $p={2,\infty}$. However, there are attacks that work outside of this traditional scheme.

CornerSearch~\cite{croce2019sparse} proposes a black-box attack based on finding an adversarial example with respect to the $l_{0}$ norm. Abandoning norm based constraints completely, Patch Attack~\cite{yang2020patchattack} replaces a certain area of the image with an adversarial patch. Likewise, ColorFool~\cite{colorfool} disregards norms and instead recolors the image to make it adversarial. While the non-traditional norm category is not strictly defined, it gives us a concise grouping that highlights the advances being made outside of the $l_{2}$ and $l_{\infty}$ based black-box attacks.    

\begin{table}[]
\centering
\begin{tabular}{|l|c|c|}
\hline
\multicolumn{3}{|c|}{Score   based Attacks} \\ \hline
\multicolumn{1}{|c|}{Attack Name} & Date & Author \\ \hline
qMeta & \textit{6-Jun-19} & Du et al.~\cite{Du2020Query-efficient}  \\ \hline
P-RGF & \textit{17-Jun-19} & Cheng et al.~\cite{cheng2019improving} \\ \hline
ZO-ADMM & \textit{26-Jul-19} & Zhao et al.~\cite{zhao2019design} \\ \hline
TREMBA & \textit{17-Nov-19} & Huang et al.~\cite{huang2019black} \\ \hline
Square & \textit{29-Nov-19} & Andriushchenko et al.~\cite{andriushchenko2020square} \\ \hline
ZO-NGD & \textit{18-Feb-20} & Zhao et al.~\cite{zhao2020towards} \\ \hline
PPBA & \textit{8-May-20} & Liu et al.~\cite{li2020projection} \\ \hline
\multicolumn{3}{|c|}{Decision based   Attacks} \\ \hline
Attack Name & \multicolumn{1}{l|}{Date} & \multicolumn{1}{l|}{Author} \\ \hline
qFool & \textit{26-Mar-19} & Liu et al.~\cite{qfool} \\ \hline
HSJA & \textit{3-Apr-19} & Chen et al.~\cite{chen2020hopskipjumpattack} \\ \hline
GeoDA & \textit{13-Mar-20} & Rahmati et al.~\cite{rahmati2020geoda} \\ \hline
QEBA & \textit{28-May-20} & Li et al.~\cite{qeba} \\ \hline
RayS & \textit{23-Jun-20} & Chen et al.~\cite{chen2020rays} \\ \hline
SurFree & \textit{25-Nov-20} & Maho et al.~\cite{surfree} \\ \hline
NonLinear-BA & \textit{25-Feb-21} & Li et al.~\cite{nonlinear_ba} \\ \hline
\multicolumn{3}{|c|}{Transfer based   Attacks} \\ \hline
\multicolumn{1}{|c|}{Attack Name} & Date & Author \\ \hline
Adaptive & \textit{3-Oct-19} & Mahmood et al.~\cite{mahmood2019buzz} \\ \hline
DaST & \textit{28-Mar-20} & Zhou et al.~\cite{dast} \\ \hline
PO-TI & \textit{13-Jun-20} & Li et al.~\cite{towards_transfer} \\ \hline
\multicolumn{3}{|c|}{Non-traditional   Attacks} \\ \hline
\multicolumn{1}{|c|}{Attack Name} & Date & Author \\ \hline
CornerSearch & \textit{11-Sep-19} & Croce et al.~\cite{croce2019sparse} \\ \hline
ColorFool & \multicolumn{1}{l|}{\textit{25-Nov-19}} & Shamsabadi et al.~\cite{colorfool} \\ \hline
Patch & \textit{12-Apr-20} & Yang et al.~\cite{yang2020patchattack} \\ \hline
\end{tabular}
\caption{Attacks covered in this survey, their corresponding attack categorization, publication date (when the first e-print was released) and author. \label{tbl:mainTable}}
\end{table}

\subsection{Paper Organization and Major Contributions}

In this paper we survey state-of-the-art black-box attacks that have recently been published. We provide three major contributions in this regard:

\begin{enumerate}
    \item \textbf{In-Depth Survey:} We summarize and distill the knowledge from 20 recent significant black-box adversarial machine learning papers. For every paper, we include explanation of the mathematics necessary to conduct the attacks and describe the corresponding adversarial model. We also provide an experimental section that brings together the results from all 20 papers, reported on three datasets (MNIST, CIFAR-10 and ImageNet).
    \item \textbf{Attack Categorization:} We organize the attacks into four different categories based on the underlying adversarial model used in each attack. We present this organization so the reader can clearly see where advances are being made under each of the four adversarial threat models. Our break down concisely helps new researchers interpret the rapidly evolving field of black-box adversarial machine learning. 
    \item \textbf{Attack Analysis Framework:} We analyze how the attack success rate is computed based on different adversarial models and their corresponding constraints. Based on this analysis, we develop an intuitive way to define the threat model used to compute the attack success rate. Using this framework, it can clearly be seen when attack results reported in different papers can be compared, and when such evaluations are invalid.  
\end{enumerate}

The rest of our paper is organized as follows: in Section~\ref{sec:scoreBasedAttack}, we summarize score based attacks. In Section~\ref{sec:decisionAttack}, we cover the papers that propose new decision based attacks. In Section~\ref{sec:transfer}, we discuss transfer attacks. The last type of attack, non-traditional attacks are described in Section~\ref{sec:NonTraditional}. After covering all the new attacks, we turn our attention to analyzing the attack success rate in Section~\ref{sec:TheoryAttackRate}. Based on this analysis, we compile the experimental results for all the attacks in Section~\ref{sec:exp}, and give the corresponding threat model developed from our new adversarial model framework. Finally, we offer concluding remarks in Section~\ref{sec:conclusion}.
\section{Score based Attacks}
\label{sec:scoreBasedAttack}
In this section we summarize recent advances in adversarial machine learning with respect to attacks that are score based or logit based. The adversarial model for these attacks allow the attacker to query the defense with input $x$ and receive the corresponding probability outputs $p_{1}(x),...,p_{k}(x)$, where $k$ is the number of classes. We also include logit based black-box attacks in this section. The logits are the pre-softmax outputs from the model, $l_{1}(x),...,l_{k}(x)$.

We cover 7 recently proposed score type attacks. These attacks include the square attack~\cite{andriushchenko2020square}, the Zeroth-Order Natural Gradient Descent attack (ZO-NGD)~\cite{zhao2020towards}, the Projection and Policy Driven Attack (PPBA)~\cite{li2020projection}, the Zeroth-order Optimization Alternating Direction Method of Multiplers (ZO-ADMM) attack~\cite{zhao2019design}, the prior-guided random gradient-free (P-RGF) attack~\cite{cheng2019improving}, the   TRansferable   EMbedding   based   Black-box   Attack (TREMBA)~\cite{huang2019black} and the qMeta attack~\cite{Du2020Query-efficient}.

\subsection{Square Attack}
The Square attack is a score based, black-box adversarial attack proposed in~\cite{andriushchenko2020square} that focuses primarily on being query efficient while maintaining a high attack success rate. The novelty of the attack comes in the usage of square shaped image perturbations which have a particularly strong impact on the predicted outputs of CNNs. This works in tandem with the implementation of the randomized search optimization protocol. The protocol is independent of model gradients and greedily adds squares to the current image perturbation if they lead to an increase in the target model's error. The attack solves the following optimization problem:
\begin{equation}
    \begin{array}{c cc}
        \underset{\hat{x}\in [0,1]^d}{\text{min}} L(f(\hat{x}), y),  &  
        \text{s.t.} & \|\hat{x}-x\|_p \leq \epsilon
    \end{array}
\end{equation}
Where $f$ is the classifier function, $K$ is the number of classes, $\hat{x}$ is the adversarial input, $x$ is the clean input, $y$ is the ground truth label, and $\epsilon$ is the maximum perturbation.
\begin{equation}
\label{losssquare}
\begin{array}{c c}
    \text{Untargeted}: &  L(f(\hat{x}),y) = f_y(\hat{x}) - \text{max}_{k \neq y}f_k(\hat{x})\\ \\
    \text{Targeted}:  & 
     L(f(\hat{x}),t) = -f_t(\hat{x})+\text{log}(\sum_{i=1}^{K}e^{f_i(\hat{x})}) 
\end{array}
\end{equation}
The attack algorithm begins by first applying random noise to the clean image. Then an image perturbation, $\delta$, is generated according to a perturbation generating algorithm defined by the attacker. If $L(f(\hat{x} + \delta), y) < L(f(\hat{x}), y)$ $\delta$ is applied to the current $\hat{x}$. This step is done iteratively until the targeted model outputs the desired label or until the max number of iterations are reached.

The distributions used for the iterative and initial image perturbations are chosen by the attacker. In~\cite{andriushchenko2020square} two different initial and iterative perturbation algorithms algorithms are proposed for the $l_2$ and $l_\infty$ norm attacks.

For the $l_\infty$ norm the perturbation is initialized by applying one pixel wide vertical stripes to the clean image. The color of each stripe is sampled uniformly from $\{-\epsilon, \epsilon\}^c$ where c is the number of color channels. The distribution used in the iterative step generates a square of a given size at a random location such that the magnitude of the perturbation in each color channel is chosen randomly from $\{-2\epsilon, 2\epsilon\}$. The resulting, clipped adversarial image will then differ from the clean image by either  $\epsilon$ or $-\epsilon$ at each modified point.

The $l_2$ norm attack is initialized by generating a grid-like tiling of squares on the clean image. The perturbation is then rescaled to have $l_2$ norm $\epsilon$ and is clipped to $[0,1]^d$. The iterative perturbation is motivated by the realization that classifiers are particularly susceptible to large, localized perturbations rather than smaller, more sparse ones. Thus the iterative attack places two squares of opposite sign either vertically or horizontally in line with each other, where each square has a large magnitude at its center that swiftly drops off but never reaches zero. After each iteration of the attack the current $x_{adv}$ is clipped such that $\|x_{adx} - x\|_p < \epsilon$ and $x_{adv} \in [0,1]^d$, where d is the dimensionality of the clean image.

The attack is tested on contemporary models like ResNet-50, Inception v3, and VGG-16-BN which are trained on ImageNet. It achieves a lower attack failure rate while requiring significantly less queries to complete than attacks like Bandits, Parsimonious, DFO-MCA, and SignHunter. Similarly the square attack is compared to the white box Projected Gradient Descent (PGD) attacks on the MNIST and CIFAR-10 datasets where it performs similarly to PGD in terms of attack success rate despite operating within a more difficult threat model.

\subsection{Zeroth-Order Natural Gradient Descent Attack}

The Zeroth-Order Natural Gradient Descent (ZO-NGD) attack is a score-based, black box attack proposed in~\cite{zhao2020towards} as a query efficient attack utilizing a novel attack optimization technique. In particular the attack approximates a Fisher information matrix over the distribution of inputs and subsequent outputs of the classifier. The attack solves the following optimization problem:
\begin{equation}
    \underset{\delta}{\text{min}} \; f(x+\delta, t), \; \|\delta\|_\infty \leq \epsilon
\end{equation}
\begin{equation}
    f(x + \delta, t) = \text{max}\{\text{log} \; p(t|x+\delta) - \underset{i \neq t}{\text{max}}\{\text{log} \; p(i|x+\delta))\}, -k\}
\end{equation}
Where $x$ is the clean image, $\delta$ is an image perturbation, $\epsilon$ is the maximum allowed image perturbation, t is the clean image's ground truth label, $p(i|x)$ is the classifier's predicted score for class $i$ given input $x$, and $f$ is the attack's loss. The attack is an iterative algorithm that initializes the image perturbation, $\delta$, as a matrix of all zeros. At each step the algorithm first approximates the gradient of the loss function, $f$, according to the following equation:
\begin{equation}
\label{zogddeltaf}
    \widehat{\nabla}f(\delta) = \frac{1}{R}\sum^R_{j=1} \frac{f(\delta + \mu u_j, t) - f(\delta, t)}{\mu}u_j
\end{equation}
Where each $u_j \sim N(0, I_d)$ is a random perturbation chosen i.i.d. from the unit sphere, $\mu$ is a smoothing parameter, and $R$ is a hyper parameter for the number of queries used in the approximation. Next, the attack approximates the gradient of the log-likelihood function. This is necessary for calculating the Fisher information matrix and subsequently the perturbation update. 
\begin{multline}
    \widehat{\nabla}\text{log} \; p(t|x+\delta) = \frac{1}{R\mu}  \sum^R_{j=1}(\text{log}\;p(t|x+\delta+\mu u_j) \\ - \text{log} \;p(t|x + \delta))u_j
\end{multline}
Here the notation is consistent with the notation seen in Equation~\ref{zogddeltaf}. This can be calculated using the same queries that were used in Equation~\ref{zogddeltaf}. The Fisher information matrix is approximated and $\delta$ is updated according to the following equations: 
\begin{equation}
    \widehat{F} = \widehat{\nabla}\text{log}\;p(t|x+\delta) \widehat{\nabla}\text{log}\;p(t|x+\delta)^T + \gamma I
\end{equation}
\begin{equation}
    \label{deltafinal}
    \delta_{k+1} = \prod (\delta_k - \lambda \widehat{F}^{-1}\widehat{\nabla}f(\delta_k))
\end{equation}
Where $\gamma$ is a constant and $\lambda$ is the attack learning rate. $\prod$ is the projection function which projects its input onto the set $S = \{\delta \mid \; (x + \delta) \in [0,1]^d , \; \|\delta\|_\infty \leq \epsilon\}$. It is also worth recognizing that $\delta$ is represented as a matrix since images, like $x$, are also represented as matrices. This makes the addition seen in Equation~\ref{deltafinal} valid. The iterative process can be continued for a predetermined number of iterations or until the perturbation yields a satisfactory result. The Fisher information matrix is a powerful tool, however its size can prove it impractical for use on datasets with larger inputs, thus an approximation of $\delta_{k+1}$ may be necessary.

The attack is tested on the MNIST, CIFAR-10, and ImageNet datasets where it achieves a similar attack success rate to the ZOO, Bandits, and NES-PGD attacks while requiring less queries to be successful. The attack is then also shown to have an extremely high attack success rate within 1200 queries on all three aforementioned datasets.

\subsection{Projection and Probability Driven Attack}
 The Projection and Probability-driven Black-box Attack (PPBA) proposed in~\cite{li2020projection} is a score based, black box attack that achieves high attack success rates while being query efficient. It achieves this by shrinking the solution space of possible adversarial inputs to those which contain low-frequency perturbations. This is motivated by an observation that contemporary neural networks are particularly susceptible to low frequency perturbations. The attack solves the following optimization problem:
 \begin{equation}
     \underset{\delta}{\text{min}} \; L(\delta) = [f(x+\delta)_t - \underset{j \neq t}{\text{max}} \; f(x+\delta)_j]^+
 \end{equation}
 Where $f(x)_j$ is the model's predicted probability that the input is of class $j$, $x$ is the clean image, $t$ is the ground truth label, $\delta$ is the adversarial perturbation, and $[\cdot]^+$ is shorthand for $\text{max}(\cdot, 0)$. The attack utilizes a sensing matrix, $A$, which is composed of a Discrete Cosine Transform matrix, $\Psi$, and a Measurement matrix, $\Phi$, along with the corresponding measurement vector, $z$. The exact design of the measurement matrix varies according to practice \cite{7096434}~\cite{7185392}. The relationship between all these variables is as follows: $A = \Psi \Phi$, $z = A\delta$, $\delta \approx A^Tz$. 
 
 
 One point to note is that $\Phi$ should be an orthonormal matrix which allows $\delta \approx A^Tz$ to be true. Once $A$ is calculated the attack utilizes a query efficient version of the random walk algorithm. In particular, the attack stores a Confusion matrix $C_j$ for each dimension $j$ of $\Delta z$, which is the change in $z$ at each iteration. $C_j$ can be seen below:
 \[
     \begin{array}{c c c c}
        & -\rho & 0  & \rho \\ \hline
        \text{\# effective steps} & e_{-\rho} & e_0 & e_\rho \\
        \text{\# ineffective steps} & i_{-\rho} & i_0 & i_\rho
     \end{array}
 \]
 Where $\rho$ is a predefined step size, $e_v$ is the number of times the loss function descended when $\Delta z_j = v$, and $i_v$ is the number of times the loss function increased or remained the same when $\Delta z_j = v$ for $v \in \{-\rho, \, 0, \, \rho\}$. The algorithm then uses $C$ to determine its sampling probability for $\Delta z_j$ as seen below:
 \begin{equation}
     P(a|\Delta z_j = v) = \frac{e_v}{e_v + i_v},\; v \in \{-\rho, 0, \rho\}
 \end{equation}
 \begin{equation}
 \label{PPDB}
     P(\Delta z_j = v) = \frac{P(\text{a}|\Delta z_j = v)}{\sum_u P(\text{a}|\Delta z_j = u)}, \; u,v \in \{-\rho, 0, \rho\}
 \end{equation}
 Where $a$ is a probabilistic variable that is true when the step is determined to be effective. The attack algorithm begins by first calculating $A$ and then initializing all values of $C$ to be 1. The iterative part of the algorithm then begins, at each step the algorithm generates a new $\Delta z$ according to the probability distribution described in Equation~\ref{PPDB}. If $L(A^T(z+\Delta z)) < L(A^Tz)$ then $z$ is updated as $z = \text{clip}(z + \Delta z)$. Here the clip function forces $x + z$ to remain within the clean image's input space, $[0,1]^d$. If at any point the perturbation generated causes the model to output an incorrect class label the attack terminates and returns the penultimate perturbation.
 
 
 
 PPBA is tested on the ImageNet dataset with the classifiers ResNet50, Inception v3 and VGG-16. PPBA achieves high attack success rates while maintaining a low query count. It is also tested on Google Cloud Vision API where it achieves a high attack success rate in this more realistic setting.
 
 \subsection{Alternating Direction Method of Multiplers Based Black-Box Attacks}

A new black-box attack framework is proposed in~\cite{zhao2019design} based on the distributed convex optimization technique, the Alternating Direction Method of Multiplers (ADMM). The advantage of using the ADMM technique is that it can be directly combined with the zeroth-order optimization attack (ZOO-ADMM) or Bayesian optimization (BO-ADMM) to create a query-efficient, gradient free black-box attack. The attack can be run with score based or decision based output from the defense. 

The main concept presented in~\cite{zhao2019design} is the conversion of the black-box attack optimization problem from a traditional constrained optimization problem, into an unconstrained objective function that can be iteratively solved using ADMM. The original formulation of the black-box attack optimization problem can be written as:

\begin{equation}
\label{eqADMMOrig}
\begin{aligned}
& \underset{\delta}{\text{minimize}}
& & f(x_{0}+\delta, t) + \gamma D(\delta) \\
& \text{subject to}
& & (x_{0} + \delta) \in [0,1]^{d}, \|\delta\|_{\infty} \leq \epsilon
\end{aligned}
\end{equation}
where $f(\cdot)$ is the loss function of the classifier, $\delta$ is the perturbation added to the original input $x_{0}$, $t$ is the target class that the adversarial example $(x_{0}+\delta)$ should be misclassified as and $D$ is a distortion function to limit the difference between the adversarial example and $x_{0}$. In Equation~\ref{eqADMMOrig}, $\gamma$ controls the weight given to the distortion function and $\epsilon$ specifies the maximum tolerated perturbation.

Instead of directly solving Equation~\ref{eqADMMOrig}, the constraints can be moved into the objective function and an auxiliary variable $z$ can be introduced in order to write the optimization problem in an ADMM style form:

\begin{equation}
\label{eqADMMObjFun}
\begin{aligned}
& \underset{\delta, z}{\text{minimize}}
& & f(x_{0}+\delta, t) + \gamma D(\delta) + \mathcal{I}(z) \\
& \text{subject to}
& & z = \delta
\end{aligned}
\end{equation}
where $\mathcal{I}(z)$ is $0$ if $(x_{0}+z) \in [0,1]^{d}, \|z\|_{\infty} \leq \epsilon$ and $\infty$ otherwise. The augmented Lagrangian of Equation~\ref{eqADMMObjFun} is written as:

\begin{equation}
\label{eqADMMLang}
\begin{aligned}
 & \mathcal{L}(z, \delta, u) = \gamma D(z) + \mathcal{I}(z) + f(x_{0} + \delta, t) 
 & & \\
 & +\frac{\rho}{2} \|z-\delta+\frac{1}{\rho}u\| -\frac{1}{2\rho}\|u\|_{2}^{2}
 \end{aligned}
\end{equation}
where $u$ is the Lagrangian multiplier and $\rho$ is a pentalty parameter. Equation~\ref{eqADMMLang} can be iteratively solved using ADMM in the $k^{th}$ step through the following update equations: 


\begin{equation}
\label{EqSolveZ}
z^{k+1} = \underset{z}{\text{arg min}} \mathcal{L}(z, \delta^{k}, u^{k})
\end{equation}

\begin{equation}
\label{EqSolveDelta}
\delta^{k+1} = \underset{\delta}{\text{arg min}} \mathcal{L}(z^{k+1}, \delta, u^{k})
\end{equation}

\begin{equation}
\label{EqSolveU}
u^{k+1} = u^{k} + \rho(z^{k+1}-\delta^{k+1})
\end{equation}
While Equation~\ref{EqSolveZ} has a closed form solution, minimizing Equation~\ref{EqSolveDelta} requires a gradient descent technique like stochastic gradient decent, as well as access to the gradient of $f(x_{0}+\delta, t)$. In the black-box setting this gradient is not available to the adversary and hence must be estimated using a special approach. If the gradient is estimated using the random gradient estimation technique, then the attack is referred to as ZOO-ADMM. Similarly, if the gradient is estimated using bayesian optimization, the attack is denoted as BO-ADMM.

The new attack framework is experimentally verified on the CIFAR-10 and MNIST datasets. The results of the paper~\cite{zhao2019design} show ZOO-ADMM outperforms both BO-ADMM and the original boundary attack presented in~\cite{brendel2018decision}. This performance improvement comes in the form of smaller distortions for the $l_{1}$, $l_{2}$ and $l_{\infty}$ threat models and in terms of less queries used for the ZOO-ADMM attack.

\subsection{Improving Black-box Adversarial Attacks with Transfer-based Prior}

Initial adversarial machine learning black-box attacks were developed based on one of two basic principles. In query based black-box attacks~\cite{brendel2018decision}, the gradient is directly estimated through querying. In transfer based attacks, the gradient is computed based on a trained model's gradient that is available to the attacker~\cite{papernot2017practical}.
In~\cite{cheng2019improving} they propose combining the query and transfer based attacks to create a more query efficient attack which they call the prior-guided random gradient-free method (P-RGF).   

The P-RGF attack is developed around accurately and efficiently estimating the gradient of the target model $f$. The original random gradient-free method~\cite{nesterov2017random} estimates the gradient as follows:

\begin{equation}
\label{RGFEq}
    \hat{g} = \frac{1}{q} \sum_{i=1}^{q} \frac{f(x+\sigma u_{i}, y)-f(x,y)}{\sigma} \cdot u_{i}
\end{equation}
where $q$ is the number of queries used in the estimate, $\sigma$ is a parameter to control the sampling variance, $x$ is the input with corresponding label $y$ and $\{u_i\}^{q}_{i=1}$ are random vectors sampled from distribution $\mathcal{P}$. It is important to note that by selecting $\{u_i\}^{q}_{i=1}$ carefully (according to priors) we can create a better estimate of $g$. In P-RGF this choice of $\{u_i\}^{q}_{i=1}$ is done by biasing the sampling using a transfer gradient $v$. The transfer gradient $v$ comes from a surrgoate model that has been independently trained on the same data as the model whose gradient is currently being estimated. In the attack it is assumed that we have white-box access to the surrogate model such that $v$ is known. 

The overall derivation of the rest of the attack from ~\cite{cheng2019improving} goes as follows: first we discuss the appropriate loss function $L(\cdot)$ for $\hat{g}$. We then discuss how to pick $\{u_i\}^{q}_{i=1}$ such that $L(\cdot)$ is minimized. To determine how closely $\hat{g}$ (the estimated gradient) follows $g$ (the true model gradient) the following loss function is used~\cite{cheng2019improving}: 


\begin{equation}
\label{RGFLoss}
\begin{aligned}
& \underset{b \geq 0 }{\text{min}}
& & \mathbb{E} \| \nabla_{x} f(x) -b\hat{g} \|_{2}^{2}
\end{aligned}
\end{equation}
where $b$ is a scaling factor included to compensate for the change in magnitude caused by $\hat{g}$ and the expectation is taken over the randomness of the estimation algorithm. For notational convenience we write $\nabla_{x} f(x)$ as $\nabla f(x)$ in the remainder of this subsection. It can be proven that if $x$ is differentiable at $f$ then the loss function given in Equation~\ref{RGFLoss} can be expressed as:
\begin{equation}
\label{RGFLoss2}
\begin{aligned}
& \underset{\sigma \rightarrow 0 }{\text{lim}} L(\hat{g}) = 
& & \| \nabla f(x) \|_{2}^{2}-\frac{H(\textbf{C},x)^{2}}{(1-\frac{1}{q})H(\textbf{C}^{2},x)+\frac{1}{q}H(\textbf{C},x)^{2})}\\
\end{aligned}
\end{equation}
where $H(\textbf{C},x)=\nabla f(x)^{T} \textbf{C} \nabla f(x)$ and $\textbf{C}=\mathbb{E}[u_{i} u_{i}^{T}]$. Through careful choice of $\textbf{C}$, $L(\hat{g})$ can be minimized to accurately estimate the gradient, thereby making the attack query efficient. $\textbf{C}$ can be decomposed in terms of the transfer gradient $v$ as: 
\begin{equation}
\label{Prior10}
    \textbf{C} = \lambda v v^{T} + \frac{1-\lambda}{D-1}(\textbf{I}-vv^{T})
\end{equation}
where $\{\lambda_{i}\}^{D}_{i=1}$ and $\{v_{i}\}^{D}_{i=1}$ are the eigenvalues and orthonormal eigenvectors of $\textbf{C}$. 
To exploit the gradient information of the transfer model, $u_{i}$ is then randomly generated in terms of $v$ to satisfy Equation~\ref{Prior10}:
\begin{equation}
\label{FinalUPrior}
u_{i}=\sqrt{\lambda} \cdot v + \sqrt{1-\lambda} \cdot \overline{(\textbf{I}-vv^{T})\xi_{i}}
\end{equation}
where $\lambda$ controls the magnitude of the transfer gradient $v$ and $\xi_{i}$ is a random variable sampled uniformly from the unit hypersphere. 

The overall P-RGF method for estimating the gradient $g$ is as follows: First $\alpha$, the cosine similarity between the transfer gradient $v$ and the model gradient $g$ is estimated through a specialized query based algorithm~\cite{cheng2019improving}. Next $\lambda$ is computed as a function of $\alpha$, $q$ and the input dimension size $D$. Note we omitted the $\lambda$ equation and explanation in our summary for brevity. After computing $\lambda$, the estimate of the gradient $\hat{g}$ is iteratively done $Q$ times in a two step process. In the first step of the $q^{th}$ iteration, $u_{q}$ is generated using Equation~\ref{FinalUPrior}. In the second step $\hat{g}$ is calculated as: $\hat{g} = \hat{g} + \frac{f(x+\sigma u_{q}, y)-f(x,y)}{\sigma} \cdot u_{q}$, where $q$ denotes the $q^{th}$ iteration. After $Q$ iterations have been complete, the final gradient estimate is given as $\hat{g} \leftarrow \frac{1}{Q} \hat{g}$.

The P-RGF attack is tested on ImageNet. The surrogate model to get the transfer gradient in the attack is set as ResNet-152. Attacks are done on different ImageNet CNNs which include Inception v3, VGG-16 and ResNet50. The P-RGF attack outperforms other completing techniques in terms of having a higher attack success rate and lower number of queries for most networks. 


\subsection{Black-Box Adversarial Attack with Transferable Model-based Embedding}

The TRansferable EMbedding based
Black-box Attack (TREMBA)~\cite{huang2019black} is an attack that uniquely combines transfer and query based black-box attacks. In conventionally query based black-box attacks, the adversarial image is modified by iteratively fine tuning the noise that is directly added to the pixels of the original image. In TREMBA, instead of directly altering the noise, the embedding space of a pre-trained model is modified. Once the embedding space is modified, this is translated into noise for the adversarial image. The advantage of this approach is that by using the pre-trained model's embedding as a search space, the amount of queries needed for the attack can be reduced and the attack efficiency can be increased. 

The attack generates the perturbation $\delta$ for input $x$ using a generator network $\mathcal{G}$. The generator network is comprised of two components, an encoder $\mathcal{E}$ and a decoder $\mathcal{D}$. The encoder maps $x$ to $z$, a latent space i.e., $z=\mathcal{E}(x)$. The decoder $\mathcal{D}$ takes $z$ as input. The outputs of the decoder $\mathcal{D}$ is used to compute the  perturbation $\delta$ which is defined as $\delta = \epsilon \text{tanh}(\mathcal{D}(z))$. The tanh function is used to normalize the output of the decoder $\mathcal{D}(z)$ between $-1$ and $1$ such that the final adversarial perturbation $\delta$ is bounded i.e. $||\delta||_{\infty} \leq \epsilon$.  

To begin the untargeted version of the attack, the generator network $\mathcal{G}$ is first trained. For an individual sample $(x_{i},y_{i})$, we denote the probability score associated with the correct class label during training as:
\begin{equation}
\label{eq:TrueP}
    P_{true}(x_{i},y_{i}) = F_{s}(\epsilon \cdot \text{tanh}(\mathcal{G}(x_{i}))+x_{i}))_{y_{i}}
\end{equation}
where $\epsilon$ is the maximum allowed perturbation, $\mathcal{G}(\cdot)$ is the output from the generator and $F_{s}(\cdot)_{i}$ is the $i^{th}$ component of the output vector of the source model $F_{s}$. In this attack formulation the adversary is assumed to have white-box access to a pre-trained source model $F_{s}$ which is different from the target model under attack. The incorrect class label with the maximum probability during training is:
\begin{equation}
\label{eq:falseP}
\begin{aligned}
& P_{false}(x_{i},y_{i})=
& \underset{j \neq y_{i}}{\text{max}}
& & F_{s}(\epsilon \cdot \text{tanh}(\mathcal{G}(x_{i}))+x_{i}))_{j}
\end{aligned}
\end{equation}
Using Equation~\ref{eq:TrueP} and Equation~\ref{eq:falseP} the loss function for training the generator for an untargeted attack is given as:
\begin{equation}
    \mathcal{L}_{untarget}(x_{i},y_{i}) = \text{max}(P_{true}(x_{i},y_{i})-P_{false}(x_{i},y_{i}), -\kappa)
\end{equation}
where $(x_{i},y_{i})$ are individual training samples in the training dataset and $\kappa$ is a transferability parameter (higher $\kappa$ makes the adversarial examples more transferable to other models~\cite{carlini2017towards}).

Once $\mathcal{G}$ is trained the perturbation $\delta$ can be calculated as a function of the embedding space $z$. The embedding space $z$ is iteratively computed:
\begin{equation}
    z_{t} = z_{t-1} - \frac{\eta}{b} \sum^b_{i=1} \mathcal{L}_{untarget} \nabla_{z_{t-1}} \text{log}(\mathcal{N}(v_{i}|z_{t-1}, \sigma^{2})) 
\end{equation}
where $t$ is the iteration number, $\eta$ is the learning rate, $b$ is the sample size, $v_{i}$ is a sample from the gaussian distribution $\mathcal{N}(z_{t-1},\sigma^2)$ and  $\nabla_{z_{t-1}}$ is the gradient of $z_{t}$ 
estimated using the Natural Evolution Strategy (NES)~\cite{ilyas2018black}.

Experimentally TREMBA is tested on both the MNIST and ImageNet datasets. The attack is also tested on the Google Cloud Vision API. In general, TREMBA achieves a higher attack success rate and uses less queries for MNIST and ImageNet, as compared to other attack methods. These other attack methods compared in this work include P-RGF, NES and AutoZOOM.

\subsection{Query-Efficient Meta Attack}
In the query-efficient meta attack~\cite{Du2020Query-efficient}, high query-efficiency is achieved through the use of meta-learning to observe previous attack patterns. This prior information is then leveraged to infer new attack patterns through a reduced number of queries.
First, a meta attacker is trained to extract information from the gradients of various models, given specific input, with the goal being to infer the gradient of a new target model using few queries. That is, an image $\textbf{x}$ is input to models $\mathcal{M}_1,...,\mathcal{M}_n$ and a max-margin logit classification loss is used to calculate losses $l_1,...,l_n$ as follows:
\begin{equation} \label{mm_loss}
l_i(\textbf{x}) = \text{max }[\text{log}[\mathcal{M}_i(\textbf{x})]_t - \underset{j \neq t}{\text{ max }} \text{log}[\mathcal{M}_i(\textbf{x})]_j,0]
\end{equation}
where $t$ is the true label, $j$ is the index of other classes, $[\mathcal{M}_i(\textbf{x})]_t$ is the probability score produced by the model $\mathcal{M}_i$, and $[\mathcal{M}_i(\textbf{x})]_j$ refers to the probability scores of the subsequent classes. 

After one step back-propagation is performed, $n$ training groups for the universal meta attacker are assembled, consisting of input images $\mathbb{X} = \{ \textbf{x} \}$ and gradients $\mathbb{G}_i = \{ \textbf{g}_i\}, i = 1,...,n $ where $\textbf{g}_i = \nabla_{\textbf{x}} l_i(\textbf{x})$.
In each training iteration, $K$ samples are drawn from a task $\mathcal{T}_i = (\mathbb{X},\mathbb{G}_i)$. For meta attacker model $\mathcal{A}$ with parameters $\boldsymbol{\theta}$, the updated parameters $\boldsymbol{\theta}^{'}$ are computed as: $\boldsymbol{\theta}_i^{'} := \boldsymbol{\theta} - \alpha \nabla_{\boldsymbol{\theta}} \mathcal{L}_i(\mathcal{A}_{\boldsymbol{\theta}})$, where $\mathcal{L}_i$ is the loss corresponding to task $\mathcal{T}_i$.

The meta attack parameters are optimized by incorporating $\boldsymbol{\theta}_i^{'}$ across all tasks $\{ \mathcal{T}_i \}_i=1,...,n $ according to:
\begin{equation} \label{task}
\boldsymbol{\theta} := \boldsymbol{\theta} + \epsilon \frac{1}{n} \sum_{i=1}^{n} (\boldsymbol{\theta}_i^{'} - \boldsymbol{\theta})
\end{equation}
The training loss of this meta attacker $\mathcal{A}_{\boldsymbol{\theta}}$ employs mean-squared error, as given below:
\begin{equation} \label{mse1}
\mathcal{L}_i (\mathcal{A}_{\boldsymbol{\theta}} ) = \rVert \mathcal{A}_{\boldsymbol{\theta}} (\mathbb{X}_s) - \mathbb{G}_i^s \lVert_2^2
\end{equation}
where the set $(\mathbb{X}_s, \mathbb{G}_i^s)$ refers to the $K$ samples selected for training from $(\mathbb{X}, \mathbb{G}_i)$ for $\boldsymbol{\theta}$ to $\boldsymbol{\theta}_i^{'}$ .

The high-level objective of such a meta attacker model $\mathcal{A}$ is to produce a helpful gradient map for attacking that is adaptable to the gradient distribution of the target model. To accomplish this efficiently, a subsection $q$ of the total $p$ gradient map coordinates are used to fine-tune $\mathcal{A}$ every $m$ iterations ~\cite{Du2020Query-efficient}, where $q \ll p$. In this manner, $\mathcal{A}$ is trained to be able to produce the gradient distribution of various input images and learns to predict the gradient from only a few samples through this selective fine-tuning. It is of importance to note that query efficiency is further reinforced by performing the typically query-intensive zeroth-order gradient estimation only every $m$ iterations. 

Empirical results on MNIST, CIFAR-10, and tiny-ImageNet attain comparable attack success rates to other untargeted black-box attacks. However, the attack significantly outperforms prior attacks in terms of the number of queries required in the targeted setting~\cite{Du2020Query-efficient}.

\section{Decision based Attacks}
\label{sec:decisionAttack}
In this section, we discuss recent developments in adversarial machine learning with respect to attacks that are decision based. The adversarial model for these attacks allows the attacker to query the defense with input $x$ and receive the defense's final predicted output. In contrast to score based attacks, the attacker does not receive any probabilistic or logit outputs from the defense.

We cover 7 recently proposed decision based attacks. These attacks include the Geometric decision-based attack ~\cite{rahmati2020geoda}, Hop Skip Jump Attack ~\cite{chen2020hopskipjumpattack}, RayS Attack ~\cite{chen2020rays}, Nonlinear Black-Box Attack ~\cite{nonlinear_ba}, Query-Efficient Boundary-Based Black-box Attack ~\cite{qeba}, SurFree attack ~\cite{surfree}, and the qFool attack ~\cite{qfool}.

\subsection{Geometric Decision-based Attacks}
Geometric decision-based attacks (GeoDA) are a subset of decision based black box attacks proposed in~\cite{rahmati2020geoda} that can achieve high attack success rates while requiring a small number of queries. The attack exploits a low mean curvature in the decision boundary of most contemporary classifiers within the proximity of a data point. In particular the attack uses a hyperplane to approximate the decision boundary in the vicinity of a data point to effectively find the local normal vector of the decision boundary. The normal vector can then be used to modify the clean image in such a way that the model outputs an incorrect class label. Thus the attack solves the following optimization problem:
\begin{equation}
\begin{array}{c c}
    \underset{v}{\text{min}} & \|v\|_p \\
    \text{s.t.} & w^T(x+v) - w^Tx_B = 0
\end{array}
\end{equation}
Where $w$ is a normal vector to the decision boundary, and $x_B$ is point on the decision boundary and close to the clean image, $x$. $x_B$ can be found by adding random noise, $r$, to $x$ until the classifier's predicted label changes, then performing a binary search in the direction of $r$ to get $x_B$ as close to the decision boundary as possible:
\begin{equation}
\begin{array}{c}
    x_B = x + \underset{r}{\text{min}}\|r\|_2  \\
    \text{ s.t. } \hat{k}(x_B) \neq \hat{k}(x)
\end{array}
\end{equation}

Where $\hat{k}(\cdot)$ returns the top-1 label of the target classifier. The normal vector to the decision boundary is found in the following way: $N$ image perturbations, $\eta_i$, are randomly drawn from a multi-variate normal distribution $\eta_i \sim  \mathcal{N}(0, \Sigma)$ \cite{liu2019geometry}. The model is then queried on the top-1 label of each $x_B + \eta_i$ where $x_b$ is a boundary point close to the clean image, $x$. Each $\eta_i$ is then classified as follows: 
\begin{equation}
    \mathcal{S}_{adv} = \{\eta_i \mid \hat{k}(x_b + \eta_i) \neq \hat{k}(x)\}
\end{equation}
\begin{equation}
    \mathcal{S}_{clean} = \{\eta_i \mid \hat{k}(x_b + \eta_i) = \hat{k}(x)\}
\end{equation}

From here the normal vector to the decision boundary can then be estimated as:
\begin{equation}
    \hat{w}_N = \frac{\bar{\mu}_N}{\|\bar{\mu}_N\|_2}
\end{equation}

\begin{equation}
    \begin{array}{c c}
    \text{where} &  \bar{\mu}_N = \frac{1}{N}\sum_{i=1}^N\rho_i\eta_N\\ \\
    \text{and} & \rho_i = \left\{\begin{array}{c c}
        1 & \eta_i \in S_{adv}\\
        -1 & \eta_i \in S_{clean}
    \end{array}   \right.
\end{array}
\end{equation}
Finally the image can be modified using the following update:
\begin{equation}
    x_{adv} = x + \hat{r}\hat{w}_N
\end{equation}
\begin{equation}
    \begin{array}{c c}
    \text{where} &  \hat{r} = \text{min}\{r>0 \mid \hat{k}(x+rv) \neq \hat{k}(x)\}\\ \\
    \text{and} & v = \frac{1}{\|\hat{w}_N\|_a}\odot \text{sign}(\hat{w})
\end{array}
\end{equation}

Here $\odot$ refers to the point-wise product and $a = {\frac{p}{p-1}}$. This process is done iteratively, at each iteration the previous iteration's $x_{adv}$ is used to calculate $\hat{w}$ which is then added to the original $x$ to find the current iteration's $x_{adv}$ as seen above. 

The attack is experimentally tested on the ImageNet dataset. The experiments show GeoDA outperforms the Hop Skip Jump Attack, Boundary Attack, and qFool by producing smaller image perturbations and requiring less iterations, and thus less queries, to complete. 

\subsection{Hop Skip Jump Attack}
The Hop Skip Jump Attack (HSJA) is a decision based, black-box attack proposed in~\cite{chen2020hopskipjumpattack} that achieves both a high attack success rate and a low number of queries. The attack is an improvement on the previously developed Boundary Attack \cite{brendel2017decision} in that it implements gradient estimation techniques at the edge of a model's decision boundary in order to more efficiently create adversarial inputs to the classifier. Similarly to many other adversarial attacks, HSJA attempts to change the predicted class label of a given input, $x$, while minimizing the perturbation applied to the input. Thus the following optimization problem is proposed:
\begin{equation}
\begin{array}{c cc}
    \underset{x'}{\text{min}} \; d(x', x^*) &  \text{s.t.} & \phi_{x^*}(x') = 1
\end{array}
\end{equation}
\begin{equation}
    \phi_{x^*}(x') = \text{sign}(S_{x^*}(x'))
\end{equation}
\begin{equation}
\label{hsjas}
    S_{x^*}(x') = \left\{\begin{array}{cc}
            \underset{c \neq c^*}{\text{max}}\;F_c(x') - F_{c^*}(x') &  \text{(Untargeted)}\\
            F_{c^\dagger}(x') - \underset{c \neq c^\dagger}{\text{max}}\;F_c(x') & \text{(Targeted)} \end{array} \right.
\end{equation}
Here $F_c$ is the predicted probability of class $c$, $x'$ is the adversarial input, $x^*$ is the clean input, and $d$ is a distance metric. This unique optimization formulation allows HSJA to approximate the gradient of Equation~\ref{hsjas} and thus more accurately and efficiently solve the optimization problem.

The attack algorithm starts by adding random noise, $\delta$, to the clean image, $x^*$, until the model's predicted class label changes to the desired label. Once a desired random perturbation is found the iterative process is initiated and $x^* + \delta$ is stored in $x_0$ which becomes an iterative parameter written as $x_t$ for step number $t$. From here a binary search is performed to find the decision boundary between $x^*$ and $x_t$. At the decision boundary the following operation is used to approximate the gradient of the decision boundary: 
\begin{equation}
    \widehat{\Delta S}(x_t, \delta_t) = \frac{1}{1-B}\sum^{B}_{b=1}(\phi_{x^*}(x_t+\delta_t u_b) - \overline{\phi_{x^*}})u_b
\end{equation}
\begin{equation}
    \overline{\phi_{x^*}} = \frac{1}{B} \sum^{B}_{b=1} \phi_{x^*}(x_t+\delta_t u_b)
\end{equation}
Where $\delta_t = d_t^{-1}\|x_{t-1}-x^*\|_p$ and $d_0 = \|x_0 - x^*\|$ is a small, positive parameter. Each $u_b$ is randomly drawn i.i.d. from the uniform distribution over the d-dimensional sphere. The additional term, $\overline{\phi_{x^*}}$, is used to attempt to mitigate the bias induced into the estimation by $\delta$. Once the gradient of the decision boundary is found an update direction is found using the following formulation:
\begin{equation}
    v_t(x_t, \delta_t) = \left\{ \begin{array}{cl}
        \widehat{\Delta S}(x_t, \delta_t) / \|\widehat{\Delta S}(x_t, \delta_t)\|_2 &  \mbox{if } p = 2\\
        \text{sign}(\widehat{\Delta S}(x_t, \delta_t)) & \mbox{if } p = \infty 
    \end{array} \right.
\end{equation}
Once this update direction is found a step size must be determined. The step size is initialized as $\xi_t = \|x_t - x^*\|_p/\sqrt{t}$ and is halved until $\phi_{x^*}(x_t + \xi_t v_t) \neq 0$. Then $x_t$ is updated by $x_t = x_t + \xi_t v_t$ and $d_t$ is updated by $d_t = \|x_t - x^*\|_p$. This process is continued for a predetermined $T$ iterations.

In~\cite{chen2020hopskipjumpattack} HSJA is tested on the MNIST, CIFAR-10, CIFAR-100, and ImageNet datasets. HSJA outperforms the Boundary Attack and Opt Attack in terms of median perturbation magnitude and attack success rate. HSJA is also tested against multiple defenses on the MNIST dataset, where it performs better than Boundary Attack and Opt Attack when all attacks are given an equal number of queries.

\subsection{RayS Attack}

The RayS attack is a query efficient, decision based, black-box attack proposed in~\cite{chen2020rays} as an alternative to zeroth-order gradient attacks. The attack employs an efficient search algorithm to find the nearest decision boundary that requires less queries then other contemporary decision based attacks while maintaining a high attack success rate. Specifically, the attack formulation turns the continuous problem of finding the closest decision boundary into a discrete optimization problem:
\begin{equation}
    \underset{d \in \{-1,1\}^n}{\text{min}}\; g(d) = \underset{r}{\text{arg min}}\; \mathbbm{1}\{f(x+\frac{rd}{\|d\|_2}) \neq y\}
\end{equation}
Where $x$ is the clean sample which is assumed to be a vector without loss of generality, $y$ is the ground truth label of the clean sample, $f$ is the classifier's prediction function, $d$ is a direction vector determining the direction of the perturbation in the input space, $r$ is a scalar projected onto $d$ determining the magnitude of the perturbation, and $n$ is the dimensionality of the input. This converts the continuous problem of finding the direction to the closest decision boundary into a discrete optimization problem over $d \in \{-1,1\}^n$ which contains $2^n$ possible options. 


The attack algorithm finds a direction, $d$, and a radius, $r$, in the input space as its final output for the attack. They can then be converted into a perturbation by projecting $r$ onto $d$. The attack begins by choosing some initial direction vector, $d$, and setting $r = \infty$. The iterative process comes in multiple stages, $s$, where at each stage $d$ is cut into $2^s$ equal and uniformly placed blocks. The algorithm then iterates through each of these blocks, swapping the sign of each value in the current block at a given iteration and storing the modified $d$ into $d_{temp}$. If $f(x + r\cdot d_{temp}) = y$ the algorithm skips searching $d_{temp}$ as it requires a larger perturbation than $d$ to change the classifier's predicted label. If $f(x + r\cdot d_{temp}) \neq y$ the algorithm performs a binary search in the direction of $d_{temp}$ to find the smallest $r$ such that $f(x + r\cdot d_{temp}) \neq y$ remains true. Finally $d$ is updated to $d_{temp}$ and $r$ is updated to be the smallest radius found in the binary search. 

The RayS attack is experimentally tested in~\cite{chen2020rays} on the MNIST, CIFAR-10, and ImageNet datasets. It outperforms other black-box attacks like HSJA and SignOPT in terms of both average number of queries and attack success rate on the MNIST and CIFAR-10 datasets. On the ImageNet dataset, HSJA achieves a lower number of average queries than RayS, but attains a significantly lower attack success rate. The RayS attack is also compared to white box attacks like Projected Gradient Descent (PGD) where it outperforms the attack on the MNIST and CIFAR-10 datasets, in terms of the attack success rate.

\subsection{Nonlinear Projection Based Gradient Estimation for Query Efficient Blackbox Attacks}
\label{nonlinearba}
The Nonlinear Black-box Attack (NonLinear-BA) is a query efficient, nonlinear gradient projection-based boundary blackbox attack~\cite{nonlinear_ba}. This attack innovatively overcomes the gradient inaccessibility of blackbox attacks by utilizing vector projection for gradient estimation. AE, VAE, and GAN are used to perform efficient projection-based gradient estimation. ~\cite{nonlinear_ba} shows that NonLinear-BA can outperform the corresponding linear projections of HSJA and QEBA, as NonLinear-BA provides a higher lower bound of cosine similarity between the estimated and true gradients of the target model.
\\ \indent There are three components of NonLinear-BA: the first is gradient estimation at the target model’s decision boundary. While high-dimensional gradient estimation is computationally expensive, requiring numerous queries ~\cite{nonlinear_ba}, projecting the gradient to lower dimensional supports greatly improves the estimation efficiency of NonLinear-BA. This desired low dimensionality is achieved through the latent space representations of generative models, e.g., AE, VAE, and GAN. 

The gradient projection function $\textbf{f}$ is defined as $\textbf{f} : \mathbb{R}^n \rightarrow \mathbb{R}^m$, which maps the lower-dimensional representative space $\mathbb{R}^n$ to the original, high-dimensional space $\mathbb{R}^m$, where $n \leq m$. The sample unit latent vectors $v_b$'s in $\mathbb{R}^n$ are randomly sampled to generate the perturbation vectors $u_b = \textbf{f}(v_b) \in \mathbb{R}^m$. 

Thus, the gradient estimator is as follows:
\begin{equation}
\label{ba_1}
\widetilde{\nabla S}(x^{(t)}_{adv}) = \frac{1}{B} \sum_{b=1}^{B} \text{sgn}(S(x^{(t)}_{adv} + \delta \textbf{f}(v_b)))\textbf{f}(v_b)
\end{equation}
where $\widetilde{\nabla S}$ is the estimated gradient, $x_{adv}$ is the boundary image at iteration t, $S$ is the difference function that indicates whether the image has been successfully perturbed from the original label to the malicious label, the function $\text{sgn}(S(\cdot))$ denotes the sign of this difference function, and $\delta$ is the size of the random perturbation to control the gradient estimation error.

The second component of NonLinear-BA is moving the boundary-image $x_{adv}$ along the estimated gradient direction:
\begin{equation}
\label{ba_2}
\hat{x}_{t+1} = x^{(t)}_{adv} + \xi_t \cdot \frac{\widetilde{\nabla S}}{ {\lVert \widetilde{\nabla S} \rVert}_2}
\end{equation}
where $\xi_t$ is a step size chosen by searching with queries.

Finally, in order to enable the gradient estimation in the next iteration and move closer to the target image, the adversarial image $x_{adv}$ is mapped back to the decision boundary through binary search. This search is aided by queries which seek to find a fitting weight $\alpha_t$:
\begin{equation}
\label{ba_3}
x^{(t+1)}_{adv} = \alpha_t \cdot x_{tgt} + (1-\alpha_t) \cdot \hat{x}_{t+1}
\end{equation}
where $x_{tgt}$ is the target image, i.e., the original image whose correct label $x_{adv}$ seeks to achieve with a crafted perturbed image.

TNonLinear-BA is evaluated on both offline model ImageNet, CelebA, CIFAR10 and MNIST datasets, as well as commercial online APIs. The nonlinear projection-based gradient estimation black-box attacks achieve better performance compared with the state-of-the-art baselines. The authors in~\cite{nonlinear_ba} discover that when the gradient patterns are more complex, the NonLinear-BA-GAN method fails to keep reducing the MSE after a relatively small number of queries and converges to a poor local optima.

\subsection{QEBA: Query-Efficient Boundary-Based Blackbox Attack}
Black-box attacks can be query-free or query-based. Query-free attacks are transferability based; query access is not required, as this type of attack assumes the attacker has access to the training data such that a substitute model may be constructed. Query-based attacks can be further categorized into score-based or boundary-based attacks. In a score-based attack, the attacker can access the class probabilities of the model. In a boundary-based attack, only the final model prediction label, rather than the set of prediction confidence scores, is made accessible to the attacker. Both score-based and boundary-based attacks require a substantial number of queries.

\indent One challenge of reducing the number of queries needed for a boundary-based attack is that it is difficult to explore the decision boundary of high-dimensional data without making many queries. The Query-Efficient Boundary-based Blackbox Attack (QEBA) seeks to reduce the queries needed by generating queries through adding perturbations to an image~\cite{qeba}. Thus, probing the decision boundary is reduced to searching a smaller, representative subspace for each generated query. Three representative subspaces are studied by ~\cite{qeba}: spatial transformed subspace, low frequency subspace, and intrinsic component subspace. The optimality analysis of gradient estimation query efficiency in these subspaces is shown in ~\cite{qeba}.

\indent QEBA performs an iterative algorithm comprised of three steps: first, estimate the gradient at the decision boundary, which is based on the given representative subspace, second, move along the estimated gradient, and third, project to the decision boundary with the goal of moving towards the target adversarial image. These steps follow the same mathematical details as given in Equation~\ref{ba_1} to~\ref{ba_3} in Section~\ref{nonlinearba}. Representative subspace optimizations from spatial, frequency, and intrinsic component perspectives are then consequently explored; these subspace-based gradient estimations are shown to be optimal as compared to estimation over the original space~\cite{qeba}.

Results for the attack are provided for models trained on ImageNet and models trained on the CelebA dataset. The results show the MSE vs the number of queries, indicating that the three proposed query efficient methods outperform HSJA significantly. The authors also show that the proposed QEBA significantly reduces the required number of queries. In addition, the attack yields high quality adversarial examples against both offline models (i.e. ImageNet) and online real-world APIs such as Face++ and Azure.

\subsection{SurFree: a Fast Surrogate-free Blackbox Attack}
\label{surfree}
Many black-box attacks rely on substitution, i.e., a surrogate model is used in place of the target model, the aim being that adversarial examples crafted to attack this surrogate model will effectively transfer to the target classifier. Accordingly, an accurate gradient estimate to create the substitute model requires a substantial number of queries. 

\indent By contrast, SurFree is a geometry-based black-box attack that does not query for a gradient estimate~\cite{surfree}. Instead, SurFree assumes that the boundary is a hyperplane and exploits subsequent geometric properties as follows. Consider the pre-trained classifier to be $f : [0,1]^D \rightarrow \mathbb{R}^C$. A given input image $\textbf{x}$ produces the label
\begin{math}
cl(\textbf{x}) := \text{arg max}_k \text{\;}  f_k(\textbf{x}),
\end{math}
where $f_k(\textbf{x})$ is the predicted probability of class class $k, 1 \leq k \leq C$. The goal of an untargeted attack is to find an adversarial image $\textbf{x}_a$ that is similar to a classified image $\textbf{x}_o$ such that $cl(\textbf{x}_a) \neq cl(\textbf{x}_o)$. Thus, an outside region is defined as 
\begin{math}
\mathcal{O} := \{ \textbf{x} \in \mathbb{R}^D : cl(\textbf{x}) \neq cl( \textbf{x}_o ) \}
\end{math} 
The desired, optimal adversarial image is then:
\begin{equation} \label{surf1}
\textbf{x}^*_a = \underset{\textbf{x} \in \mathcal{O}}{\text{arg min }} ||\textbf{x} - \textbf{x}_o||
\end{equation}
A key assumption of SurFree is that if a point $\textbf{y} \in \mathcal{O}$, then there exists a point $\textbf{x}_b \in \overline{x_o y}$ which can be found that lies on the boundary, denoted as $\partial \mathcal{O}$. Further, it is assumed that the boundary $\partial \mathcal{O}$ is an affine hyperplane that passes through $\textbf{x}_{b,1}$ in $\mathbb{R}^D$ with normal vector $\textbf{N}$. Considering a random basis with span $(\textbf{x}_{b,1} - \textbf{x}_o)^\perp$ composed of $D - 1$ vectors $\{ {\textbf{v}_i} \}^{D-1}_{i = 1}$, the inner product between $\textbf{N}$ and $(\textbf{x}_{b,k} - \textbf{x}_o) \propto \textbf{u}_k$ can be iteratively increased by:
\begin{equation} \label{surf2}
\textbf{N}^\top \textbf{u}_k = \prod_{i = 1}^{D-k} \cos (\psi_{D - i})
\end{equation}
where $\textbf{u}_k$ is the vector that spans the plane containing $\textbf{x}_o$, and $\textbf{x}_{b,D} \in \mathcal{O}$ and $(\textbf{x}_{b,D} - \textbf{x}_o)$ is colinear with $\textbf{N}$, which points to the projection of $\textbf{x}_o$ along the boundary of the hyperplane.


Additionally, restricting perturbations to a low dimensional subspace improve the estimation of the projected gradient. The low dimensional subspace is carefully chosen to incorporate meaningful, prior information about the visual content of the image. This further aids in implementing a low query budget.

\indent It is experimentally shown that SurFree bests state-of-the-art techniques for limited query amounts (e.g., one thousand queries) while attaining competitive results in unlimited query scenarios ~\cite{surfree}. The geometric details of approximating a hyperplane surrounding a boundary point are left to ~\cite{surfree}.

The authors present attack results using the criteria of number of queries, and the resulting distortion on the attacked image, on the MNIST and ImageNet datasets. SurFree drops significantly faster than other compared attacks (QEBA and GeoDA) to lower distortions (most notably from 1 to 750 queries.

\subsection{A Geometry-Inspired Decision-Based Attack}
qFool is a decision-based attack that requires few queries for both non-targeted and targeted attacks ~\cite{qfool}. qFool relies on exploiting the locally flat decision boundary around adversarial examples. In the non-targeted attack case, the gradient direction of the decision boundary is estimated based upon the top-1 label result of each query. An adversarial example is then sought in the estimated direction from the original image. In the targeted attack case, gradient estimations are made iteratively from multiple boundary points from a starting target image. Query efficiency is further improved by seeking perturbations in low-dimensional subspace.

Prior literature ~\cite{fawzi_ref_for_qfool} has shown that the decision boundary has only a small curvature near the presence of adversarial examples. This observation is thus exploited by ~\cite{qfool} to compute an adversarial perturbation $v$. It conceptually follows that the direction of the smallest adversarial perturbation $v$ for the input sample $x_0$ is the gradient direction of the decision boundary at $x_{adv}$. Due to the blackbox nature of attack, this gradient cannot be computed directly; however, from the knowledge that the boundary is relatively flat, the classifier gradient at point $x_{adv}$ will be nearly identical to the gradient of other neighboring points along the boundary. Therefore, the direction of $v$ can be suitably approximated by $\xi$, the gradient estimated at a neighbor point $P$. Thus, an adversarial example $x_{adv}$ from $x_0$ is sought along $\xi$.

The three components of the untargeted qFool attack involve an initial point, gradient estimation, and a directional search. To begin with, the original image $x_0$ is perturbed by a small, random Gaussian noise to produce a starting point $\mathcal{P}$ on the boundary:
\begin{equation} \label{geo1}
\mathcal{P} := x_0 + \underset{r}{\text{min }} {\lVert r \rVert}_2 \text{ s.t. } f_\theta (\mathcal{P}) \neq f_\theta (x_0), r \sim \mathcal{N}(0,\sigma)
\end{equation}

Noise continues to be added ($\mathcal{P} = x_0 + r_j$) until the image is misclassified. Next, the top-1 label of the classifier is used to estimate the gradient of the boundary $\nabla f(\mathcal{P})$:
\begin{equation} \label{geo2}
z_i = 
\begin{cases}
& -1 \;\;\;\; f(\mathcal{P} + \nu_i) = f(x_0) \\
& +1 \;\;\;\; f(\mathcal{P} + \nu_i) \neq f(x_0)
\end{cases}
, i = 1, 2,...,n
\end{equation}

where $\nu_i$ are randomly generated vectors with the same norm to perturb $\mathcal{P}$ and $f(\mathcal{P} + \nu_i)$ is the label produced by querying the classifier.

For the final step of qFool, the gradient direction at point $x_{adv}$ can be approximated by the gradient direction at point $\mathcal{P}$, i.e., $\nabla f(\mathcal{P}) \approx \xi$. The adversarial example $x_{adv}$ can thus be found by perturbing the decision boundary in the direction of $\xi$ until the decision boundary is reached. Using binary search, this costs only a few queries to the classifier.

For a targeted attack, the objective becomes perturbing the input image to be classified as a particular target class, i.e., $f_\theta (x_0 + v) = t$ for a target class $t$. Thus, the starting point of this attack is selected to be an arbitrary image $x_t$ that belongs to the target class $t$. Due to the potentially large distance between $x_0$ and $x_t$, the assumption of a flat decision boundary between the initial and targeted adversarial regions no longer holds. Instead, a linear interpolation in the direction of ($x_t - x_0$) is utilized to find a starting point $\mathcal{P}_0$:
\begin{equation} \label{geo3}
\mathcal{P}_0 := \underset{\alpha}{\text{min }} (x_0 + \alpha \cdot \frac{x_t - x_0}{{\lVert x_t - x_0 \rVert}_2} ) \text{ s.t. } f_\theta (\mathcal{P}_0) = t
\end{equation}

The gradient direction estimation of $\xi_0$ at $\mathcal{P}_0$ follows the same method as outlined for untargeted attacks.

The qFool attack is experimentally demonstrated on the ImageNet dataset by attacking VGG-19, ResNet50 and Inception v3. The results show that qFool is able to achieve a smaller distortion in terms of MSE, as compared to the Boundary Attack when both attacks use the same number of queries. However, the overall attack success rate for qFool is not reported. The authors also test qFool on the Google Cloud Vision API. 
\section{Transfer Attacks}
\label{sec:transfer}
In this section, we explore recent advances in adversarial machine learning with respect to transfer attacks. The adversarial model for these attacks allows the attacker to query the target defense and or access some of the target defense's training dataset. The attacker then uses this information to create a synthetic model which the attacker then attacks using a white box attack. The adversarial inputs generated from the white box attack on the synthetic model are then transferred to the targeted defense.

We cover 3 recently proposed transfer attacks. These attacks include the Adaptive Black-Box Transfer attack~\cite{mahmood2019buzz}, DaST attack~\cite{dast} and the Transferable Targeted attack~\cite{towards_transfer}.

\subsection{The Adaptive Black-box Attack}

A new transfer based black-box attack is developed in~\cite{mahmood2019buzz} that is an extension of the original Papernot attack proposed in~\cite{papernot2017practical}. Under this threat model the adversary has access to the training dataset $(X,Y)$, and query access to the classifier under attack, $C$. In the original Papernot formulation of the attack, the attacker labels the training data to create a new training dataset $(X, C(X))$. The adversary is then able to train synthetic model $S$ on $(X, C(X))$ while iteratively augmenting the dataset using a synthetic data generation technique. This results in a trained synthetic model $S(w_{s})$. In the final step of the attack, a white-box attack generation method $\phi(\cdot)$ is used in conjunction with the trained synthetic model $S$ in order to create adversarial examples $X_{adv}$:
\begin{equation}
    X_{adv} = \phi(X_{clean}, S, w_{s})
\end{equation}
where $X_{clean}$ are clean testing examples and $\phi$ is a white-box attack method i.e. FGSM~\cite{goodfellow2014explaining}.

The enhanced version of the Papernot attack is called the mixed~\cite{mahmood2019buzz} or adaptive black-box attack~\cite{mahmood2020beware}. Where as in the original Papernot attack $0.3\%$ of the training data is used, the adaptive version increases the strength of the adversary by using anywhere from $1\%$ to $100\%$ of the original training data. Beyond this, the attack generation method $\phi$ is varied to account for newer white-box attack generation methods that have better transferability. In general the most effective version of the attack replaces $\phi_{\text{FGSM}}$ with $\phi_{\text{MIM}}$, the Momentum Iterative Method (MIM)~\cite{MIMdong2018boosting}. The MIM attack computes an accumulated gradient~\cite{MIMdong2018boosting}:

\begin{equation}
\label{eq:mim}
g_{t+1} = \mu \cdot g_t + \frac{J(x^{adv}_{t},y)}{||\nabla_x J(x^{adv}_{t},y)||_{1}}
\end{equation}
where $J(\cdot)$ is the loss function, $\mu$ is the decay factor and $x^{adv}_{t}$ is the adversarial sample at attack iteration $t$. For a $L_{\infty}$ bounded attack, the adversarial example at iteration $t$ is:
\begin{equation}
\label{eq:mim_fgsm}
x^{adv}_{t+1} = x^{adv}_{t} + \frac{\epsilon}{T} \cdot \text{sign}(g_{t+1})
\end{equation}
where $T$ represents the total number of iterations in the attack and $\epsilon$ represents the maximum allowed perturbation.

In~\cite{mahmood2019buzz}, the attack is tested using the CIFAR-10 and Fashion-MNIST datasets. The adaptive black-box attack is shown to be effective against vanilla (undefended) networks, as well as a variety of adversarial machine learning defenses.  

\subsection{DaST: Data-free Substitute Training for Adversarial Attacks}
As described in the SurFree attack in~\ref{surfree}, substitute models can be difficult or unrealistic to obtain, particularly if a substantial amount of real data labeled by the target model is needed. DaST is a data-free substitute training method that utilizes generative adversarial networks (GANs) to train substitute models without the use of real data ~\cite{dast}. To address the potentially uneven distribution of GAN-produced samples, a multi-branch architecture and label-control loss for the GAN model is employed.
\\ \indent To describe the necessary context for DaST, let $\textbf{X}$ denote samples from the target model $T$, $\bar{y}$ and $y'$ denote the true labels and target labels of the samples $\textbf{X}$, respectively, and let $T(y|\textbf{X},\theta)$ denote the target model parameterized by $\theta$. Then, the objective of a targeted attack becomes:

\begin{equation} \label{dast1}
\begin{split}
\min_{\epsilon}||\epsilon|| \text{ subject to} & \underset{y_i}{\text{ argmax }} T(y_i | \overline{\textbf{X}} = \textbf{X} + \epsilon, \theta ) = y' \\
     & \text{and } ||\epsilon|| \leq r
\end{split}
\end{equation}

where $\epsilon$ and $r$ are the sample and upper bounds of the perturbation, respectively, and $\overline{\textbf{X}} = \textbf{X} + \epsilon$ refer to the adversarial examples that lead the target model $T$ to misclassify a sample with a selected wrong label.

\indent To further provide adequate context for DaST, a white-box attack under these settings would have full access to the gradient construction of the target model $T$ and thus leverage this information to generate adversarial examples. In a black-box substitute attack under these settings, a substitute model $\hat{T}$ would stand-in for the target model, and the adversarial examples generated to attack $\hat{T}$ would then be transferred to attack $T$. Thus, coming to the settings of a data-free black-box substitute attack, DaST utilizes a GAN to synthesize a training set for $\hat{T}$ that is as similar as possible to the training set of the target model $T$. 
\\ \indent To this end, the substitute training set crafted by the GAN aims to be evenly distributed across all categories of labels, which are produced from $T$. To accomplish this, for $N$ categories, the generative network in ~\cite{dast} is designed to contain $N$ upsampling deconvolutional components, which then share a post-processing convolutional network. The generative model $G$ randomly samples a noise vector $\textbf{z}$ from the input space as well as the variable label $n$. $\textbf{z}$ then enters the $n$-th upsampling deconvolutional network and the shared convolutional network to produce the adversarial sample $\hat{\textbf{X}} = G(\textbf{z},n)$. The label-control loss for $G$ is given as: 
\begin{equation} \label{dast2}
\mathcal{L}_c = \text{CE}(T(G(\textbf{z},n)),n)
\end{equation}
where CE is the cross-entropy.

To approximate the gradient information of $T$ to train a label-controllable generative model, the following objective function is used:
\begin{equation} \label{dast3}
\underset{\text{D}}{\text{min}} \; d(T(\hat{\textbf{X}}),D(\hat{\textbf{X}}))
\end{equation}

For the same inputs, the outputs of $D$ will approach the outputs of $T$ for the same inputs as training proceeds. Thus, $D$ replaces $T$ in Equation~\ref{dast2}:
\begin{equation} \label{dast4}
\mathcal{L}_c = \text{CE}(D(G(\textbf{z},n)),n)
\end{equation}

The loss of G is then updated  as:
\begin{equation} \label{dast5}
\mathcal{L}_G = e^{-d(T,D)} + \alpha \mathcal{L}_c
\end{equation}

where $\alpha$ is the weight of the label-control loss.

As the training stage progresses, as does the imitation quality of $D$, leading to a diverse set of synthetically generated samples labeled by $T$. These data-free substitute training-produced samples are then used to attack $T$.

DaST reduces the need for adversarial substitute attacks by utilizing GANs to generate synthetic samples, and thus can train substitute models without the requirement of any real data. Authors present results on using DaST to train a substitute model for adversarial attacks on the CIFAR-10 and MNIST trained models. The substitute models trained by DaST perform better than baseline models on FGSM and C\&W attacks (targeted).

\subsection{Towards Transferable Targeted Attack}
Crafting targeted transferable examples has the dual challenges of noise curing, i.e., the decreasing gradient magnitude in iterative attacks that results in momentum accumulation, and the difficulty of moving adversarial examples toward a target class while creating distance from the true class. To this end, Li et al ~\cite{towards_transfer} propose a novel targeted, transferable attack that applies the Poincar\'e distance to combat noise curing by creating a self-adaptive gradient, and employs metric learning to improve the distance from an adversarial example's true label.

To overcome the drawback of the Poincar\'e distance fused logits failing to satisfy ${\lVert l(x) \rVert}_2 < 1$, this attack normalizes logits by the $l_1$ distance. To overcome the problem of potential infinite distances between a point and its target label, a constant of $\xi = 0.0001$ is subtracted from the one-hot target label $y$. The Poincar\'e distance metric loss is given as: 
\begin{equation} \label{targ1}
\mathcal{L}_{P_o}(x,y) = d(u,v) = \text{arccosh}(1+ \delta(u,v))
\end{equation}
where $d$ refers to the Poincar\'e distance, $l_k(x)$ indicates the output logits of the $k$-th model, $u = l_k(x) / {\lVert l_k (x) \rVert}_1, v = \text{max} \{y - \xi,0 \}$, and $l(x)$ refer to the fused logits. 
By contrast, this attack formulates adversarial examples through the fusion of logits from a combination of models, as shown below:
\begin{equation} \label{targ2}
l(x) = \sum_{k=1}^{K} w_k l_k(x)
\end{equation}
where $K$ is the number of ensemble models, $l_k(x)$ indicates the output logits of the $k$-th model, and $w_k$ is the ensemble weight of the the $k$-th model, with $w_k > 0$, $\sum_{k=1}^{K} w_k = 1$. Note that the fused logits are the average of the ensemble models.

Triplet loss is a popular targeted attack loss function that increases the distance between the adversarial example and the true label, while decreasing the distance between the adversarial example and the target label~\cite{towards_triplet_loss_ref}. A common triplet loss function appears as:
\begin{equation} \label{trip_loss}
\mathcal{L}_{trip}(x^a, x^p, x^n) = [D(x^a, x^p) - D(x^a, x^n) + \gamma]_+
\end{equation}
where $x^a, x^p, x^n$ are the anchor, positive, and negative examples, respectively, where $x^a$ and $x^p$ are of the same class, while $x^n$ is of a different class than $x^a$. The distance $d$ is based on the embedding vector for the anchor, positive, and negative networks in the triplet configuration. Additionally, $\gamma \geq 0$ is a hyperparameter that regulates the margin between the distance metrics $D(x^a, x^p)$ and $D(x^a, x^n)$.

A drawback of standard triplet loss is the need to sample new data, which is often infeasible in a targeted attack. Instead, this work formulates the triplet input as the logits of clean images, $l(x_{clean})$, the one-hot target label, $y_{tar}$, and the true label, $y_{true}$:
\begin{multline} \label{trip_loss2}
\mathcal{L}_{trip}(y_{tar}, l(x_i), y_{true}) = [D(l(x_i), y_{tar}) \\ - D(l(x_i), y_{true}) + \gamma]_+
\end{multline}
Due to $l(x^{adv})$ not being normalized,  angular distance is used as a distance metric (note that $x_i$ corresponds to $x_{adv}$ below):
\begin{equation} \label{trip_ang}
D(l(x^{adv}), y_{tar}) = 1 - \frac{|l(x^{adv}) \cdot y_{tar}|}{ {\lVert l(x^{adv}) \rVert}_2 {\lVert y_{tar} \rVert}_2}
\end{equation}

However, the usage of angular loss does not account for the influence of the norm on the loss value. Thus, an additional triplet loss term appears in the final loss function, as shown below:
\begin{equation} \label{loss}
\mathcal{L}_{all} = \mathcal{L}_{P_o}(l(x), y_{tar}) + \lambda \cdot \mathcal{L}_{trip}(y_{tar}, l(x_i), y_{true})
\end{equation}
where $x$ is the original, clean input image and $x_i$ is the result of the $i$-th iteration of the perturbation of $x$, with the final result being $x_{adv}$.

Results on ImageNet illustrate that this attack ~\cite{towards_transfer} achieves improved success rates over traditional attacks for white and black-box models in the targeted setting.
\section{Non-Traditional Norm Attacks}
\label{sec:NonTraditional}
In this section, we discuss recent developments in black-box attacks that use non-traditional norm threat models. In all attacks covered in previous sections the focus has been on adversaries which try to create adversarial examples with respect to the $l_{2}$ or $l_{\infty}$ norms. We denote the $l_{2}$ and $l_{\infty}$ as "traditional" norms simply because that is what a majority of the literature (17 of the 20 attacks) thus far have focused on. While this is not a strictly technical definition, it gives us a convenient and simple way to categorize the different attacks.

We cover three non-traditional norm attacks in this section. The first attack we summarize is the sparse and imperceivable attack~\cite{croce2019sparse} which focuses on black-box attacks with respect to the $l_{0}$ norm. The second non-traditional norm attack we survey is Patch Attack~\cite{yang2020patchattack}. The Patch Attack is based on completely replacing small part of the original image with an adversarial generated square (patch). The last attack we cover in this section is ColorFool~\cite{colorfool}. This attack is based on manipulating the colors within the image as opposed to directly adding adversarial noise. 

\subsection{Sparse and Imperceivable Attack}
The Sparse and Imperceivable attacks proposed in\cite{croce2019sparse} are $l_0$, black-box attacks that produce adversarial inputs while minimizing the number of perturbed pixels. The attacks come in multiple forms, but the general goal and scheme remains the same. Each attack relies on having the score based output of the network to operate, and each version of the attack attempts to solve the following optimization problem:
\begin{equation}
    \begin{array}{c}
        \text{min} \; \gamma(x' - x) \\
        \text{s.t.} \; \text{arg max} \, f(x') \neq \text{arg max} \, f(x)
    \end{array}
\end{equation}
where $x$ is the clean image, $x'$ is an adversarial image, $f$ is a function returning the classifier's score vector, and $\gamma$ is a distance function defined as follows:
\begin{equation}
\label{eq:ethanA}
        \gamma(x'-x) =  \sum^d_{i=1} \underset{j}{\text{max}}\mathbbm{1}[x'_{ij} - x_{ij}] \neq 0 
\end{equation}
Where $x_{ij}$ refers to the $i^{th}$ pixel of the $j^{th}$ color channel. It is important to note that color images are typically represented as three 2-D matrices, with one matrix corresponding to each color channel. However, in the mathematical formulation for these attacks, they treat each color channel as a 1-D matrix for notational convenience. 

In Equation~\ref{eq:ethanA}, essentially $\gamma$ counts the number of pixels in the adversarial image that deviate from the original image. There are three versions of the attack: $l_0$, $l_0 + l_\infty$, $l_0 + \sigma$. Where $l_0 + l_\infty$ and $l_0 + \sigma$ add their own additional constraints on the optimization problem as outlined below:
\[
    \begin{array}{ cc}
        \text{attack type} &  \text{Additional Constraint} \\ \hline
         l_0 + l_\infty & \|x' - x\| \leq \epsilon \\
         l_0 + \sigma & \text{Perturbations must be imperceivable}
    \end{array}
\]
Where $\epsilon$ is the maximum allowed perturbation magnitude. Each of the attack variants follow the same scheme, with the main difference being the amount each pixel is perturbed. 

The attack begins by first iterating through each pixel in the image and generating a set of pixel perturbations $\{x'_{ij}\}$ according to the following equations:
\[
    \begin{array}{c c}
        \text{Attack Type} &  \text{Pixel Perturbation}\\ \hline
        l_0 & x'_{ij} \in \{0,1\} \\
        l_0 + l_\infty & x'_{ij} = x_{ij} \pm \epsilon \\
        l_0 + \sigma & x'_{ij} = (1 \pm \kappa \sigma_{ij})x_{ij}
    \end{array}
\]
Here $\sigma$ is the standard deviation of the image in color channel $j$ in proximity to pixel $x_{ij}$. It is calculated as follows:
\begin{equation}
    \sigma_{ij} \sqrt{\text{min}\{\sigma^{y}_{ij}, \sigma^{x}_{ij}\}}
\end{equation}
Where $\sigma^{x}_{ij}$ is the standard deviation of $x_{ij}$ and the two pixels adjacent to it horizontally in color channel $j$, and $\sigma^{y}_{ij}$ is the same but for the two pixels adjacent to $x_{ij}$ vertically. The $\sigma$ term is essential to what makes the $l_0 + \sigma$ attack imperceivable to humans. It allows the attack to avoid perturbations near edges in the image as they are more easily perceivable. It also focuses the attack on increasing the intensity of pixels rather than modifying their color. Each pixel perturbation is then clipped into the $[0,1]$ range. After this each generated $x'_{i,j}$ is sorted in decreasing order according to the value of the following equation:
\begin{equation}
    \pi^r(x'_{ij}) = f_r(x'_{ij}) - f_c(x'_{ij})
\end{equation}
Where $c$ is the ground truth label of $x$, $r$ is any class label other than $c$, $\pi^r(x')$ is the score of perturbation $x'$ with respect to class $r$, and $f_r$ is the model's predicted value for class $r$ in the score vector. The perturbations $x'_{i,j}$ are then sorted by their $\pi^r$ score for each $r$. 

After the sorting, an iterative process begins. At each iteration a number of maximum pixels perturbed, $k$, is chosen, starting at one pixel and progressing to $k_{max}$ by the end. During each of these iterations an inner loop iterative process begins. The inner loop iterates through each of the possible class labels, $r$, other than $c$, applying $k$ of the top $N$ single pixel perturbations with respect to class $r$ to the clean image. If at any point in the algorithm a perturbation leads to a changed class label, the algorithm stops and returns the penultimate perturbation. 
The attack is tested on the MNIST and CIFAR-10 datasets where it achieves a high attack success rate while perturbing a small amount of total pixels on average. The attack is compared to both white-box and black-box attacks where it achieves a similar attack success rate to attacks like C$\&$W attack and Sparse Fool while having the median least pixels perturbed per attack.

\subsection{Patch Attack}
 The Patch Attack is a black box attack proposed in~\cite{yang2020patchattack} that utilizes textured patches and reinforcement learning to generate adversarial images. Each patch is cut out from from images in a pre-generated library of textures. Each texture is designed such that neural networks strongly associate them with a particular class label in a given dataset. The attack is perceivable to the human eye and applies a large magnitude perturbation to the clean image. However, the perturbation is localized and can be shrunk through optimization techniques. In this attack the reinforcement learning agent solves the following optimization problem:
 \begin{equation}
    \text{min} \; L(y, y') = -r \cdot \ln (P)
 \end{equation}
Where $L$ is the attack loss, $y$ is the target model's predicted score for the ground truth class, $y' \neq y$ is the model's predicted score for a class other than $y$, $r$ is the reinforcement learning agent's reward, and $P$ is the agent's output probability of taking action $a$ that lead to the most recent reward. $r = \text{ln}(y')$ in a targeted attack, and $r = \text{ln}(y'-y)$ in an untargeted attack. The reinforcement learning agent's policy network is represented by an LSTM and a fully connected layer. An action is defined as follows:
\begin{equation}
    a = \{u_1, v_1, i, u_2, v_2\}
\end{equation}
Where $i$ is a texture index in the texture library, $u_1$ and $v_1$ are corner positions used to crop the texture, and $u_2$ and $v_2$ are corner positions denoting where the cropped texture should be placed on the clean image. The attack algorithm is iterative, at each time step the environment state is determined and an agent is trained. Once trained the agent outputs a probability distribution over the possible actions. One action is sampled from the distribution and the associated patch is applied. 

Before the attack the texture images must be obtained externally or generated by an algorithm. In the latter case the generation algorithm is initialized with a CNN trained on the target dataset. In our description of the attack, we denote this CNN as the texture CNN to distinguish it from the CNN being attacked. A set of data is also chosen to be used for the texture generation. The data is pre-processed using Grad-CAM~\cite{selvaraju2017grad} which masks out areas of the image that are irrelevant to the primary texture information. For each convolutional layer, $j$, in the $i^{th}$ block of the texture CNN, a feature map $F^j_i$ is generated. Furthermore, the corresponding Gram matrix, $G_i^j$ is also calculated for each feature activation of each image. Each $F^j_i$ and $G_i^j$ will help encode the most important texture information in input image according to the texture CNN. The computation of the feature maps and Gram matrices are described in detail in~\cite{7780634}. 

For each image, the Gram matrices generated are then flattened and concatenated into the vector $\bar{G}$ which encodes the texture information. From here, the $\bar{G}$s are organized by the original class label of the input that generated them. This is done so that the final textures can be labeled and to maximize the effectiveness of each patch. For each class label, the $\bar{G}$s are clustered into $N$ clusters. In~\cite{yang2020patchattack} $N$ is chosen to be 30 in order to have sufficient diversity in the texture pool. In practice, the best value of $N$ will vary based upon dataset and application. For each cluster, $\bar{G_c}$, is then used to generate a feature embedding of the final texture images. This is done by minimizing the following optimization problem over $G_t$:
\begin{equation}
    L = \lambda (\bar{G} - G_t)^2
\end{equation}
Where $\lambda$ is a weight constant and $G_t$ is the feature embedding of the texture image used to generate the final texture image. We omitt some details of the attack explanation for brevity, further details are given in~\cite{gatys2015texture}. 

The attack is tested on the ImageNet dataset where it achieves a high attack success rates while covering small portions of the clean image with patches. The attack also shows an ability to maintain its high attack success rate even when defense techniques are applied to the  classifier.

\subsection{ColorFool: Semantic Adversarial Colorization}
ColorFool ~\cite{colorfool} presents a content-based black-box adversarial attack with unrestricted perturbations that selectively manipulates colors within chosen ranges to thwart classifiers, while remaining undetected by humans. 
ColorFool operates on the independent $a$ and $b$ channels of the perceptually uniform Lab color space ~\cite{colorfool_27_ref}. Color modifications are implemented without changing the lightness, $L$, of the given image. Further, ColorFool solely selects perturbations within a defined natural-color range for particular acceptable categories ~\cite{colorfool_30_ref}. 

ColorFool divides images into sensitive and non-sensitive regions to be considered for color modification. Sensitive regions, defined $\mathbb{S} =  {\{\textbf{S}_k\} }^S_{k=1} $ are separated from non-sensitive regions, defined $\overline{\mathbb{S}} =  {\{\overline{\textbf{S}}_k\} }^{\overline{S}}_{k=1} $, where $\mathcal{S} = \mathbb{S} \cup \overline{\mathbb{S}}$. 

The color of the sensitive regions, $\mathbb{S}$, is modified to generate the adversarial set $\dot{\mathbb{S}}$ as follows:
\begin{equation} \label{colorfool1}
\dot{\mathbb{S}} = \{ \dot{\textbf{S}}_k : \dot{\textbf{S}}_k = \gamma(\textbf{S}_k) + \alpha[0,N^a_k,N^b_k]^T \}^S_{k=1}
\end{equation}
where color channel $a$ ranges from green (-128) to red (+127), color channel $b$ ranges from blue (-128) to yellow (+127), and the brightness $L$ ranges from black (0) to white (100) within the $Lab$ color space ~\cite{colorfool_27_ref}. Further, $\gamma(\cdot)$ converts the intensities of an $RGB$ image to the $Lab$ colorspace, and $N^a_k \in \mathcal{N}^a_k$ and $N^b_k \in \mathcal{N}^b_k$ are randomly chosen adversarial perturbations from the set of natural color ranges $\mathcal{N}^a_k$ and $\mathcal{N}^b_k$ ~\cite{colorfool_30_ref} within the $a$ and $b$ channels.

The color of the non-sensitive regions, $\overline{\mathbb{S}}$, is modified as follows to produce to the set $\dot{\overline{\mathbb{S}}}$:
\begin{equation} \label{colorfool2}
\dot{\overline{\mathbb{S}}} = \{ \dot{\overline{\textbf{S}}}_k : \dot{\overline{\textbf{S}}}_k = \gamma (\overline{\textbf{S}}_k) + \alpha[0,\overline{N}^a,\overline{N}^b]^T \}^{\overline{S}}_{k=1}
\end{equation}
where $\overline{N}^a,\overline{N}^b \in \{-127,...128\}$ are randomly chosen within the ranges of $a$ and $b$, respectively. Note that the full ranges of $a$ and $b$ are considered, as non-sensitive regions are able to undergo greater intensity changes.

The modified sensitive and non-sensitive regions are combined to generate the adversarial image $\dot{\textbf{X}}$, as shown below:
\begin{equation} \label{colorfool3}
\dot{\textbf{X}} = Q( \gamma^{-1} ( \sum_{k=1}^{S} \dot{\textbf{S}}_k + \sum_{k=1}^{\overline{S}} \dot{\overline{\textbf{S}}}_k ) )
\end{equation}
where $Q(\cdot)$ is the quantization function that keeps the generated adversarial image within the dynamic range of pixel values, i.e., $\dot{\textbf{X}} \in \mathbb{Z}^{w,h,c}$, and $\gamma^{-1}(\cdot)$ is the inverse function that converts image intensities from the $Lab$ color space to $RGB$.

ColorFool provides robustness to defenses that utilize filters, adversarial training or modified training loss functions. Additionally, ColorFool is less detectable than restricted attacks, including JPEG compression. The empirical results were presented on the Private-Places365, CIFAR-10, and ImageNet datasets, indicating higher success rates in the previously mentioned categories.

\section{Attack Success Rate Analysis}
\label{sec:TheoryAttackRate}
In this paper we compile the experimental results from many different sources and present them together in tabular form. While it may be tempting to directly compare attack success rates, here we give a theoretical analysis to show the fallacy of direct comparisons.

The definition of a \textit{successful} adversarial example varies between papers based on which constraints are enforced during the execution of the attack. We can formalize this as follows: For classifier $C$, the associated set of clean correctly identified examples is denoted as $\mathcal{X}(C)$ such that:
\begin{equation}
    \mathcal{X}(C) = \{ (x_i,y_i)\in \mathcal{X}_t \ : \ C(x_i)=y_i \}.
\end{equation}
where $\mathcal{X}_t$ is the entire set of testing images. When classifier $C$ is attacked, we can formally define the constraints on the attacker for a query-based black-box attack using the following threat vector: $W_{threat}=[w_{c}, w_{q}, w_{\epsilon}]$ where $w_{i} \in \{0,1\}$. $w_{c}=1$ corresponds to a successful attack definition where the classifier must produce the wrong class label i.e., $C(x_{adv}) \neq y_{i}$. Likewise, $w_{q}=1$ corresponds to a successful attack where the adversarial example is generated within a fixed number of queries $q_{adv}$, and the number of queries are less than the query budget $q$: $(q-q_{adv}) \geq 0$.
Lastly $w_{\epsilon}=1$ ensures that the adversarial example falls within a certain $||l||_{p}$ distance $\epsilon$ of the original example $x_{i}$: $||x_{adv}-x_{i}||_{p} \leq \epsilon$. If any of the values in $W_{threat}$ are $0$, it simply means that the corresponding condition is not used in defining a successful adversarial example.  


We denote $\phi(x_{i},y_{i})$ as the adversarial attack method used with respect to clean sample $(x_{i},y_{i})$ returned within $q_{adv}$ queries. Written explicitly $(x_{adv}, q_{adv})=\phi(x_{i},y_{i})$ and we assume $\phi(\cdot)$ to be deterministic in nature. While this assumption may not hold true for all attacks, this simplifies the notation for the theoretical attack success rate. The attack success rate $\alpha$ over the clean set $\mathcal{X}(C)$ with respect to classifier $C$ is:
\begin{equation}
\label{attackEq}
\alpha =
\frac{
\left| \left\{ \begin{array}{c}
(x_i,y_i)\in \mathcal{X}(C) : \\ 
(x_{adv},q_{adv}) = \phi(x_{i}, y_{i}) \Rightarrow \\
(C(x_{adv}) \neq y_{i} \lor w_{c}=0) \land \\ w_{q}(q-q_{adv}) \geq 0 \land w_{\epsilon}||x_{adv}-x_{i}||_{p} \leq \epsilon \end{array} \right\} \right|}{|\mathcal{X}(C)|}, 
\end{equation}
From Equation~\ref{attackEq}, it can be seen that under the most restricted threat model ($W_{threat}=[1,1,1]$), the attack must produce an adversarial example that is misclassified i.e.,  $C(x_{adv}) \neq y_{i}$, created using limited query information i.e., $q_{adv} \leq q$ and within an acceptable $||l||_{p}$ norm. Such a threat model requires specification of the attack parameters $q$, $p$ and $\epsilon$. That is $q$ the maximum allowed number of queries per sample,  $p$ the norm measurement and $\epsilon$ the maximum allowed perturbation.

Our framework for a vector defined threat model $W_{threat}$ and corresponding attack success rate $\alpha$ is useful for two reasons. First, it allows us to categorize every query based black-box attack according to the three value system. Second and most importantly, this framework allows us to see where comparisons between attack success rates reported in different papers are legitimate. We illustrate this next point with an example from the literature.

Consider the Square attack~\cite{andriushchenko2020square} and the Zeroth-Order alternating direction method of multipliers attack (ZO-ADMM)~\cite{zhao2019design}. The untargeted attack success rate of both attacks is reported with respect to an Inception v3 network trained on ImageNet. The Square attack reports an attack success rate $\alpha$ of $92.2\%$ while the ZO-ADMM reports an attack success rate of $100\%$. Using ONLY these two values without our threat model framework makes it seem like the ZO-ADMM attack is much stronger than the Square attack, as it never fails. However, let us now consider the threat models. The threat model for ZO-ADMM is: $W_{threat}=[w_{c}=1,w_{q}=0,w_{\epsilon}=0]$. The Square attack on the other hand has the following threat model: $W_{threat}=[w_{c}=1,w_{q}=0,w_{\epsilon}=1]$. Essentially the Square attack is reporting a high attack success rate under a MORE restrictive threat model where the adversarial example must be wrongly classified \textit{and} under a certain $l_{2}$ distance from the original clean example. The ZO-ADMM attack success rate is reported only on examples that are wrongly classified, a much weaker threat model.   

\begin{table}[]
\centering
\caption{Adversarial threat models used to determine the attack success rate in each paper. $w_{c}=1$ corresponds to an attack success rate where misclassificaiton (or targeted misclassification) defines a successful adversarial attack. $w_{q}=1$ corresponds to a successful adversarial attack done within a fixed query budget. $w_{\epsilon}=1$ corresponds to a successful attack when the adversarial example is within a certain perturbation bound $\epsilon$ of the clean example.}
\begin{tabular}{|l|c|}
\hline
\multicolumn{2}{|c|}{Score   based Attacks} \\ \hline
qMeta~\cite{Du2020Query-efficient} & $w_{c}=1, w_{q}=0, w_{\epsilon}=0$ \\ \hline
P-RGF~\cite{cheng2019improving} & $w_{c}=1, w_{q}=0, w_{\epsilon}=1$ \\ \hline
ZO-ADMM~\cite{zhao2019design} & $w_{c}=1, w_{q}=0, w_{\epsilon}=0$ \\ \hline
TREMBA~\cite{huang2019black} & $w_{c}=1, w_{q}=0, w_{\epsilon}=1$ \\ \hline
Square~\cite{andriushchenko2020square} & $w_c=1, w_q=0, w_\epsilon=1$ \\ \hline
ZO-NGD~\cite{zhao2019design} & $w_c=1, w_q=0, w_\epsilon=1$ \\ \hline
PPBA~\cite{li2020projection} & $w_c=1, w_q=0, w_\epsilon=1$ \\ \hline
\multicolumn{2}{|c|}{Decision based   Attacks} \\ \hline
qFool~\cite{qfool} & $w_{c}=0, w_{q}=1, w_{\epsilon}=0$ \\ \hline
HSJA~\cite{chen2020hopskipjumpattack} & $w_c=1, w_q=1, w_\epsilon=0$ \\ \hline
GeoDA~\cite{rahmati2020geoda} & $w_c=0, w_q=1, w_\epsilon=0$ \\ \hline
QEBA~\cite{qeba} & $w_{c}=1, w_{q}=1, w_{\epsilon}=1$ \\ \hline
RayS~\cite{chen2020rays} & $w_c=1, w_q=0, w_\epsilon=1$ \\ \hline
SurFree~\cite{surfree} & $w_{c}=1, w_{q}=1, w_{\epsilon}=1$ \\ \hline
NonLinear-BA~\cite{nonlinear_ba} & $w_{c}=1, w_{q}=0, w_{\epsilon}=0$ \\ \hline
\multicolumn{2}{|c|}{Transfer based   Attacks} \\ \hline
Adaptive~\cite{mahmood2019buzz} & $w_{c}=1, w_{q}=0, w_{\epsilon}=1$ \\ \hline
DaST~\cite{dast} & $w_{c}=1, w_{q}=0, w_{\epsilon}=1$ \\ \hline
PO-TI~\cite{towards_transfer} & $w_{c}=1, w_{q}=0, w_{\epsilon}=1$ \\ \hline
\multicolumn{2}{|c|}{Non-traditional   Attacks} \\ \hline
CornerSearch~\cite{croce2019sparse} & $w_c=1, w_q=0, w_\epsilon=0$ \\ \hline
ColorFool~\cite{colorfool} & $w_{c}=1, w_{q}=0, w_{\epsilon}=0$ \\ \hline
Patch~\cite{yang2020patchattack} & $w_c=1, w_q=0, w_\epsilon=0$ \\ \hline
\end{tabular}
\end{table}
\section{Experimental Results}
\label{sec:exp}

In this section, we discuss the experimental results for all of the black-box attacks. Broadly speaking there are three common datasets that are used in measuring the attack success rate of black-box attacks.

\begin{enumerate}
    \item \textbf{MNIST} - The MNIST dataset~\cite{MNISTDataset} consists of 60,000 training images and 10,000 test images. The dataset has 10 classes, each class is a different handwritten digit, 0-9. Each digit is a $28 \times 28$ grayscale image. In general, the maximum allowed perturbation $\epsilon$ for MNIST is high as compared to other dataset. For example, $\epsilon=0.2, 0.3$ as seen in table~\ref{tbl:MNIST}. This may generally be due to the fact that MNIST images can have large perturbations, while still being visually recognizable to humans.
    \item \textbf{CIFAR-10} - The CIFAR-10 dataset~\cite{cifar10ref} consists of 50,000 training images and 10,000 test images. The 10 classes in CIFAR-10 are airplane, car, bird, cat, deer, dog, frog, horse, ship and truck. Each image is $32 \times 32 \times 3$ (color images).
    \item \textbf{ImageNet} - The ImageNet dataset~\cite{imageNetRef} contains over 14 million color images that are labeled from $\approx 20,000$ categories. The images in ImageNet are color images, however the exact size of each image varies. 
\end{enumerate}
In the following subsection we break down the analyses according to the four different attack categories.

\subsection{Score based Attack Analysis}

In table~\ref{tbl:scoreImageNet}, we show the compiled results drawn across all the papers we surveyed for the score based attacks on ImageNet classifiers. We report MNIST and CIFAR-10 results in Table~\ref{tbl:MNIST} and Table~\ref{tbl:CIFAR10}. As the majority of the attacks are done with respect to the ImageNet dataset, we relegate our discussion and analysis to those results in this subsection. 

Let us consider the $l_{\infty}=0.05$ norm adversary, untargeted attack with adversarial model $w_{c}=1, w_{q}=0, w_{\epsilon}=1$. Under this adversarial threat model, three attacks have a $99\%$ or greater attack success rate (Square, p-RGF and TREMBA). While all three attacks are done on ResNet classifiers, there is a slight difference (TREMBA and P-RGF are tested on ResNet34 and the Square attack is tested on ResNet50). Aside from this difference, if we compare results, the Square attack and TREMBA are both able to achieve a remarkable double digit query count while still maintaining a $99\%$ or greater attack success rate. Square attack requires 73 queries on average while TREMBA requires 27.  

While no attack in table~\ref{tbl:scoreImageNet} uses the most restrictive threat model (i.e. $w_{c}=1, w_{q}=1, w_{\epsilon}=1$) we can see that the most common threat model is $w_{c}=1, w_{q}=0, w_{\epsilon}=1$ making comparison between attack that use this threat model and the same classifier possible. Alternatively, one attack (ZO-ADMM) uses a highly unrestricted adversarial model $w_{c}=1, w_{q}=0, w_{\epsilon}=0$ making it impossible to directly determine the fidelity of the ZO-ADMM attack in relation to other state-of-the-art attacks.

\subsection{Decision based Attack Analysis}

In Table~\ref{tbl:decisionImageNet} we give the results for all decision based attacks that were conducted on ImageNet classifiers. Likewise, results for decsion based attacks on MNIST and CIFAR-10 can be found in Table~\ref{tbl:MNIST} and Table~\ref{tbl:CIFAR10}. Due to the large number of attacks and datasets, in this subsection we specifically focus on the decision based attacks for ImageNet CNNs. 

Let us first consider the $l_{2}$ norm decision based attacks that are targeted. For this setting, when looking at the most restricted threat model $(w_{c}=1,w_{q}=1, w_{\epsilon}=1)$ we can see that SurFree gives  the best query efficient attack (on ResNet18) with a $90\%$ attack success rate using only 500 queries. However, in terms of minimal distortion ($\epsilon=0.001$), QEBA-S and NonLinear-BA can both achieve an $80\%$ attack success rate or higher with a query budget of $10,000$. Alternatively, if we consider the $l_{\infty}$ norm and an untargeted attack, it is clear the RayS attack is the best attack. RayS achieves a $98.9\%$ attack success rate on Inception v3 using an average of $748.2$ queries per sample. This same holds true for datasets like MNIST and CIFAR-10. In both cases RayS can achieve a $99\%$ or higher attack success rate. 

It is important to note that certain threat models make attack results opaque and difficult to compare. For example, the threat model $(w_{c}=0,w_{q}=1, w_{\epsilon}=0)$ is used to report attack results for GeoDA and qFool. In this case, the median distortion is considered the independent variable (i.e. the one that changes between different attacks). However, when only the median distortion is reported this does not give any information about what percent of adversarial examples are actually misclassified (which would constitute a successful attack). Reporting the median also does not give the full picture in terms of the average distortion required to create a successful adversarial example in the attack.    
\subsection{Transfer based Attacks and Non-traditional Attacks}

In Table~\ref{tbl:nt} the results for the transfer based attacks and non-traditional attacks are shown. For the transfer attacks, each attack is done under slightly different assumptions making direction comparison difficult. For example, the Adaptive attack requires all the training data to be available to the attacker, where as in DaST the attack is specifically built around not having direct access to the original training data. Overall, we can claim that the transfer based attacks, just in terms of attack success rates, are not as high as the best score based and decision based black-box attacks. For example, the Adaptive attack has a $74\%$ attack success rate on CIFAR-10 for the $l_{\infty}$ based attacker. The decision based RayS attack has a $99.8\%$ attack success rate for CIFAR-10 (again $l_{\infty}$ norm based attack). 

For the non-traditional attacks, we can see several interesting trends. First, the patch attack has an extremely high attack success rate on ImageNet (greater than $99\%$) regardless of whether the attack is targeted or untargeted. Likewise, the $l_{0}$ based CornerSearch attack can also achieve a high untargeted attack success rate (greater than $97\%$) across both MNIST and CIFAR-10 datasets. 

The only attack that performs relatively poorly (less than $50\%$ attack success rate) is ColorFool. This may partially be due to the fact that the ColorFool attack can be run in both white-box and black-box form. The ColorFool black-box reported attack results are based on a transfer style attack as opposed to a query based method. As we mentioned above, the new score and decision based attacks (which use query information) that we survey have a higher attack success rates than the new transfer based attacks for the $l_{2}$ and $l_{\infty}$ norms. Essentially, we conjecture there may still be room to improve the black-box ColorFool attack using a query based methodology.     



\begin{table*}[]
\centering
\small
\begin{tabular}{|l|c|c|l|c|l|c|c|}
\hline
Attack Name & ASR & Avg Queries & \multicolumn{1}{c|}{Norm} & Target & \multicolumn{1}{c|}{Classifier} & Adv Threat Model & Source \\ \hline
PPBA & 84.8 & 668 & $l_{2}$, $\epsilon$=5 & U & ResNet50 & $w_{c}=1, w_{q}=0, w_{\epsilon}=1$ & ~\cite{li2020projection} \\  \hline
PPBA & 65.3 & 1051 & $l_{2}$, $\epsilon$=5 & U & Inception v3 & $w_{c}=1, w_{q}=0, w_{\epsilon}=1$ & ~\cite{li2020projection}  \\ \hline
PPBA & 96.6 & 481 & $l_{\infty}$, $\epsilon$=0.05 & U & ResNet50 & $w_{c}=1, w_{q}=0, w_{\epsilon}=1$ & ~\cite{li2020projection}  \\ \hline
PPBA & 67.9 & 1026 & $l_{\infty}$, $\epsilon$=0.05 & U & Inception v3 & $w_{c}=1, w_{q}=0, w_{\epsilon}=1$ & ~\cite{li2020projection}  \\ \hline
ZO-NGD & 97 & 582 & $l_{\infty}$, $\epsilon$=0.05 & U & Inception v3 & $w_{c}=1, w_{q}=0, w_{\epsilon}=1$ & ~\cite{zhao2020towards}  \\  \hline
Square & 99.7 & 197 & $l_{\infty}$, $\epsilon$=0.05 & U & Inception v3 & $w_{c}=1, w_{q}=0, w_{\epsilon}=1$ & ~\cite{andriushchenko2020square} \\  \hline
Square & 100 & 73 & $l_{\infty}$, $\epsilon$=0.05 & U & ResNet50 & $w_{c}=1, w_{q}=0, w_{\epsilon}=1$ & ~\cite{andriushchenko2020square}  \\ \hline
Square & 92.2 & 1100 & $l_{2}$, $\epsilon$=5 & U & Inception v3 & $w_{c}=1, w_{q}=0, w_{\epsilon}=1$ & ~\cite{andriushchenko2020square}  \\ \hline
Square & 99.3 & 616 & $l_{2}$, $\epsilon$=5 & U & ResNet50 & $w_{c}=1, w_{q}=0, w_{\epsilon}=1$ & ~\cite{andriushchenko2020square}  \\ \hline
TREMBA & 100 & 27 & $l_{\infty}$, $\epsilon$=0.05 & U & Resnet34 & $w_{c}=1, w_{q}=0, w_{\epsilon}=1$ & ~\cite{huang2019black}  \\  \hline
TREMBA & 99.44 & 443 & $l_{\infty}$, $\epsilon$=0.05 & T & Resnet34 & $w_{c}=1, w_{q}=0, w_{\epsilon}=1$ & ~\cite{huang2019black}  \\ \hline
ZO-ADMM & 98 & - & $l_{2}$ & U & Inception v3 & $w_{c}=1, w_{q}=0, w_{\epsilon}=0$ & ~\cite{zhao2019design} \\ \hline
ZO-ADMM & 97 & - & $l_{2}$ & T & Inception v3 & $w_{c}=1, w_{q}=0, w_{\epsilon}=0$ & ~\cite{zhao2019design}  \\ \hline
P-RGF & 100 & 328 & $l_{\infty}$, $\epsilon$=0.05 & U & Resnet34 & $w_{c}=1, w_{q}=0, w_{\epsilon}=1$ & ~\cite{huang2019black}  \\ \hline
P-RGF & 98.05 & 5498 & $l_{\infty}$, $\epsilon$=0.05 & T & Resnet34 & $w_{c}=1, w_{q}=0, w_{\epsilon}=1$ & ~\cite{huang2019black}  \\ \hline
P-RGF & 98.1 & 745 & $l_{2}$, $\epsilon \approx $16.43 & U & Inception v3 & $w_{c}=1, w_{q}=0, w_{\epsilon}=1$ & ~\cite{cheng2019improving} \\  \hline
P-RGF & 99.6 & 452 & $l_{2}$, $\epsilon \approx $16.43 & U & ResNet50 & $w_{c}=1, w_{q}=0, w_{\epsilon}=1$ & ~\cite{cheng2019improving} \\ \hline
P-RGF & 97.3 & 812 & $l_{\infty}$, $\epsilon$=0.05 & U & Inception v3 & $w_{c}=1, w_{q}=0, w_{\epsilon}=1$ & ~\cite{cheng2019improvingEPRINTONLY}  \\ \hline
P-RGF & 99.6 & 388 & $l_{\infty}$, $\epsilon$=0.05 & U & ResNet50 & $w_{c}=1, w_{q}=0, w_{\epsilon}=1$ & ~\cite{cheng2019improvingEPRINTONLY}  \\ \hline
\end{tabular}
\caption{Score based black-box attacks on ImageNet classifiers. The corresponding success rate (ASR) and adversarial threat model are shown for each attack along with the source paper from which the results are drawn from. \label{tbl:scoreImageNet}}
\end{table*}

\begin{table*}[b!]
\centering
\small
\begin{tabular}{|l|c|c|l|c|l|l|l|}
\hline
\multicolumn{1}{|c|}{Attack   Name} & ASR & Avg Queries & \multicolumn{1}{c|}{Norm} & Target & \multicolumn{1}{c|}{Classifier} & \multicolumn{1}{c|}{Adv Threat Model} & \multicolumn{1}{c|}{Source} \\ \hline
NonLinear-BA & 80 & 10000 & $l_{2}$, $\epsilon=$0.001 & T & ResNet18 & $w_{c}=1, w_{q}=1, w_{\epsilon}=1$ & ~\cite{li2021nonlinearEPRINTONLY}  \\ \hline
SurFree & 90 & 500 & $l_{2}$, $\epsilon=$30 & T & ResNet18 & $w_{c}=1, w_{q}=1, w_{\epsilon}=1$ & ~\cite{surfree} \\ \hline
RayS & 99.8 & 574 & $l_{\infty}$, $\epsilon$=0.05 & U & ResNet50 & $w_{c}=1, w_{q}=0, w_{\epsilon}=1$ & ~\cite{chen2020rays}  \\ \hline
RayS & 98.9 & 748.2 & $l_{\infty}$, $\epsilon$=0.05 & U & Inception v3 & $w_{c}=1, w_{q}=0, w_{\epsilon}=1$ & ~\cite{chen2020rays}  \\ \hline
QEBA-S & 82 & 10000 & $l_{2}$, $\epsilon=$0.001 & T & ResNet18 & $w_{c}=1, w_{q}=1, w_{\epsilon}=1$ & ~\cite{li2021nonlinearEPRINTONLY}  \\ \hline
QEBA-S & 74 & 10000 & $l_{2}$, $\epsilon=$0.001 & T & ResNet18 & $w_{c}=1, w_{q}=1, w_{\epsilon}=1$ & ~\cite{qeba}  \\ \hline
QEBA & 71 & 500 & $l_{2}$, $\epsilon=$30 & T & ResNet18 & $w_{c}=1, w_{q}=1, w_{\epsilon}=1$ & ~\cite{surfree}  \\ \hline
GeoDA & - & 1000 & $l_{2}$=8.16 (median) & U & ResNet50 & $w_{c}=0, w_{q}=1, w_{\epsilon}=0$ & ~\cite{rahmati2020geoda} \\ \hline
GeoDA & 79 & 500 & $l_{2}$, $\epsilon=$30 & T & ResNet18 & $w_{c}=1, w_{q}=1, w_{\epsilon}=1$ & ~\cite{surfree}  \\ \hline
HSJA & 19.9 & 749.6 & $l_{\infty}$, $\epsilon$=0.05 & U & ResNet50 & $w_{c}=1, w_{q}=0, w_{\epsilon}=1$ & ~\cite{chen2020rays} \\  \hline
HSJA & 23.7 & 652.3 & $l_{\infty}$, $\epsilon$=0.05 & U & Inception v3 & $w_{c}=1, w_{q}=0, w_{\epsilon}=1$ & ~\cite{chen2020rays}  \\  \hline
HSJA & 80 & 17000 & $l_{2}$, $\epsilon=$0.001 & T & ResNet18 & $w_{c}=1, w_{q}=1, w_{\epsilon}=1$ & ~\cite{li2021nonlinearEPRINTONLY}  \\ \hline
HSJA & 84 & 20000 & $l_{2}$, $\epsilon=$0.001 & T & ResNet18 & $w_{c}=1, w_{q}=1, w_{\epsilon}=1$ &~\cite{qeba}  \\ \hline
HSJA & 56 & 500 & $l_{2}$, $\epsilon=$30 & T & ResNet18 & $w_{c}=1, w_{q}=1, w_{\epsilon}=1$ & ~\cite{surfree}  \\ \hline
qFool & - & 1000 & $l_{2}$=16.05 (median) & U & ResNet50 & $w_{c}=0, w_{q}=1, w_{\epsilon}=0$ & ~\cite{rahmati2020geoda}  \\ \hline
\end{tabular}
\caption{Decision based black-box attacks on ImageNet CNNs. The corresponding success rate (ASR) and adversarial threat model are shown for each attack along with the original source paper. \label{tbl:decisionImageNet}}
\end{table*}

\begin{table*}[]
\centering
\small
\begin{tabular}{lccclcl}
\multicolumn{7}{c}{Score Based Attacks} \\ \hline
\multicolumn{1}{|c|}{Attack Name} & \multicolumn{1}{c|}{ASR} & \multicolumn{1}{c|}{Avg Queries} & \multicolumn{1}{c|}{Adv Threat Model} & \multicolumn{1}{c|}{Norm} & \multicolumn{1}{c|}{Target} & \multicolumn{1}{c|}{Source} \\ \hline
\multicolumn{1}{|l|}{ZO-NGD} &  \multicolumn{1}{c|}{98.7} & \multicolumn{1}{c|}{523} & \multicolumn{1}{c|}{$w_{c}=1, w_{q}=0, w_{\epsilon}=1$} & \multicolumn{1}{l|}{$l_{\infty}$, $\epsilon$=0.2} & \multicolumn{1}{c|}{U} &  \multicolumn{1}{l|}{~\cite{zhao2020towards}} \\ \hline
\multicolumn{1}{|l|}{TREMBA} & \multicolumn{1}{c|}{98} & \multicolumn{1}{c|}{1064} & \multicolumn{1}{c|}{$w_{c}=1, w_{q}=0, w_{\epsilon}=1$} & \multicolumn{1}{l|}{$l_{\infty}$, $\epsilon$=0.2} & \multicolumn{1}{c|}{U} & \multicolumn{1}{l|}{~\cite{huang2019black}} \\ \hline
\multicolumn{1}{|l|}{ZO-ADMM} & \multicolumn{1}{c|}{98.3} & \multicolumn{1}{c|}{-} & \multicolumn{1}{c|}{$w_{c}=1, w_{q}=0, w_{\epsilon}=0$} & \multicolumn{1}{l|}{$l_{2}$=1.975 (avg)} & \multicolumn{1}{c|}{T} & \multicolumn{1}{l|}{~\cite{zhao2019design}} \\ \hline
\multicolumn{1}{|l|}{P-RGF} & \multicolumn{1}{c|}{68.53} & \multicolumn{1}{c|}{16135} & \multicolumn{1}{c|}{$w_{c}=1, w_{q}=0, w_{\epsilon}=1$} & \multicolumn{1}{l|}{$l_{\infty}$, $\epsilon$=0.2} & \multicolumn{1}{c|}{U} & \multicolumn{1}{l|}{~\cite{huang2019black}} \\ \hline
\multicolumn{1}{|l|}{qMeta} & \multicolumn{1}{c|}{100} & \multicolumn{1}{c|}{749} & \multicolumn{1}{c|}{$w_{c}=1, w_{q}=1, w_{\epsilon}=0$} & \multicolumn{1}{l|}{$l_{2}$} & \multicolumn{1}{c|}{U} & \multicolumn{1}{l|}{~\cite{Du2020Query-efficient}} \\ \hline
\multicolumn{1}{|l|}{qMeta} & \multicolumn{1}{c|}{100} & \multicolumn{1}{c|}{1299} & \multicolumn{1}{c|}{$w_{c}=1, w_{q}=1, w_{\epsilon}=0$} & \multicolumn{1}{l|}{$l_{2}$} & \multicolumn{1}{c|}{T} & \multicolumn{1}{l|}{~\cite{Du2020Query-efficient}} \\ \hline
\multicolumn{7}{c}{Decision Based   Attacks} \\ \hline
\multicolumn{1}{|c|}{Attack Name} & \multicolumn{1}{c|}{ASR} & \multicolumn{1}{c|}{Avg Queries} & \multicolumn{1}{c|}{Adv Threat Model} & \multicolumn{1}{c|}{Norm} & \multicolumn{1}{c|}{Target} & \multicolumn{1}{c|}{Source} \\ \hline
\multicolumn{1}{|l|}{NonLinear-BA} & \multicolumn{1}{c|}{90} & \multicolumn{1}{c|}{5000} & \multicolumn{1}{c|}{$w_{c}=1, w_{q}=1, w_{\epsilon}=1$} & \multicolumn{1}{l|}{$l_{2}$, $\epsilon$=0.005} & \multicolumn{1}{c|}{T} & \multicolumn{1}{l|}{~\cite{li2021nonlinearEPRINTONLY}} \\ \hline
\multicolumn{1}{|l|}{RayS} & \multicolumn{1}{c|}{100} & \multicolumn{1}{c|}{107} & \multicolumn{1}{c|}{$w_{c}=1, w_{q}=0, w_{\epsilon}=1$} & \multicolumn{1}{l|}{$l_{\infty}$, $\epsilon$=0.3} & \multicolumn{1}{c|}{U} & \multicolumn{1}{l|}{~\cite{chen2020rays}} \\ \hline
\multicolumn{1}{|l|}{QEBA-S} & \multicolumn{1}{c|}{87} & \multicolumn{1}{c|}{5000} & \multicolumn{1}{c|}{$w_{c}=1, w_{q}=1, w_{\epsilon}=1$} & \multicolumn{1}{l|}{$l_{2}$, $\epsilon$=0.005} & \multicolumn{1}{c|}{T} & \multicolumn{1}{l|}{~\cite{li2021nonlinearEPRINTONLY}} \\ \hline
\multicolumn{1}{|l|}{HSJA} & \multicolumn{1}{c|}{91.2} & \multicolumn{1}{c|}{161.6} & \multicolumn{1}{c|}{$w_{c}=1, w_{q}=0, w_{\epsilon}=1$} & \multicolumn{1}{l|}{$l_{\infty}$, $\epsilon$=0.3} & \multicolumn{1}{c|}{U} & \multicolumn{1}{l|}{~\cite{chen2020rays}} \\ \hline
\end{tabular}
\caption{Decision and score based black-box attacks on MNIST CNNs. The corresponding success rate (ASR) and adversarial threat model are shown for each attack along with the original source paper.\label{tbl:MNIST}}
\end{table*}

\begin{table*}[]
\centering
\small
\begin{tabular}{lccclcl}
\multicolumn{7}{c}{Score Based Attacks} \\ \hline
\multicolumn{1}{|c|}{Attack Name} & \multicolumn{1}{c|}{ASR} & \multicolumn{1}{c|}{Avg Queries} & \multicolumn{1}{c|}{Adv Threat Model} & \multicolumn{1}{c|}{Norm} & \multicolumn{1}{c|}{Target} & \multicolumn{1}{c|}{Source} \\ \hline
\multicolumn{1}{|l|}{ZO-NGD} & \multicolumn{1}{c|}{99.2} & \multicolumn{1}{c|}{131} & \multicolumn{1}{c|}{$w_{c}=1, w_{q}=0, w_{\epsilon}=1$} & \multicolumn{1}{l|}{$l_{\infty}$, $\epsilon$=0.1} & \multicolumn{1}{c|}{U} & \multicolumn{1}{c|}{~\cite{zhao2020towards}} \\ \hline
\multicolumn{1}{|l|}{ZO-ADMM} & \multicolumn{1}{c|}{98.7} & \multicolumn{1}{c|}{-} & \multicolumn{1}{c|}{$w_{c}=1, w_{q}=0, w_{\epsilon}=0$} & \multicolumn{1}{l|}{$l_{2}$=0.417 (avg)} & \multicolumn{1}{c|}{T} & \multicolumn{1}{c|}{~\cite{zhao2019design}} \\ \hline
\multicolumn{1}{|l|}{qMeta} & \multicolumn{1}{c|}{92} & \multicolumn{1}{c|}{1765} & \multicolumn{1}{c|}{$w_{c}=1, w_{q}=1, w_{\epsilon}=0$} & \multicolumn{1}{l|}{$l_{2}$} & \multicolumn{1}{c|}{U} & \multicolumn{1}{c|}{~\cite{Du2020Query-efficient}} \\ \hline
\multicolumn{1}{|l|}{qMeta} & \multicolumn{1}{c|}{93} & \multicolumn{1}{c|}{3667} & \multicolumn{1}{c|}{$w_{c}=1, w_{q}=1, w_{\epsilon}=0$} & \multicolumn{1}{l|}{$l_{2}$} & \multicolumn{1}{c|}{T} & \multicolumn{1}{c|}{~\cite{Du2020Query-efficient}} \\ \hline
\multicolumn{7}{c}{Decision Based   Attacks} \\ \hline
\multicolumn{1}{|l|}{NonLinear-BA} & \multicolumn{1}{c|}{95.00} & \multicolumn{1}{c|}{5000.00} & \multicolumn{1}{c|}{$w_{c}=1, w_{q}=1, w_{\epsilon}=1$} & \multicolumn{1}{l|}{$l_{2}$, $\epsilon$=0.0001} & \multicolumn{1}{c|}{T} & \multicolumn{1}{c|}{~\cite{li2021nonlinearEPRINTONLY}} \\ \hline
\multicolumn{1}{|l|}{RayS} & \multicolumn{1}{c|}{99.8} & \multicolumn{1}{c|}{792.8} & \multicolumn{1}{c|}{$w_{c}=1, w_{q}=0, w_{\epsilon}=1$} & \multicolumn{1}{l|}{$l_{\infty}$, $\epsilon$=0.031} & \multicolumn{1}{c|}{U} & \multicolumn{1}{c|}{~\cite{chen2020rays}} \\ \hline
\multicolumn{1}{|l|}{QEBA-S} & \multicolumn{1}{c|}{95} & \multicolumn{1}{c|}{5000.00} & \multicolumn{1}{c|}{$w_{c}=1, w_{q}=1, w_{\epsilon}=1$} & \multicolumn{1}{l|}{$l_{2}$, $\epsilon$=0.0001} & \multicolumn{1}{c|}{T} & \multicolumn{1}{c|}{~\cite{li2021nonlinearEPRINTONLY}} \\ \hline
\multicolumn{1}{|l|}{HSJA} & \multicolumn{1}{c|}{99.7} & \multicolumn{1}{c|}{1021.6} & \multicolumn{1}{c|}{$w_{c}=1, w_{q}=0, w_{\epsilon}=1$} & \multicolumn{1}{l|}{$l_{\infty}$, $\epsilon$=0.031} & \multicolumn{1}{c|}{U} & \multicolumn{1}{c|}{~\cite{chen2020rays}} \\ \hline
\end{tabular}
\caption{Decision and score based black-box attacks on CIFAR-10 CNNs. The corresponding success rate (ASR) and adversarial threat model are shown for each attack along with the original source paper. \label{tbl:CIFAR10}}
\end{table*}

\begin{table*}[]
\small
\centering
\begin{tabular}{lllcccll}
\multicolumn{8}{c}{Transfer Based Attacks} \\ \hline
\multicolumn{1}{|l|}{Attack Name} & \multicolumn{1}{l|}{Dataset} & \multicolumn{1}{l|}{ASR} & \multicolumn{1}{l|}{Avg Queries} & \multicolumn{1}{c|}{Adv Threat Model} & \multicolumn{1}{l|}{Target} & \multicolumn{1}{l|}{Vanilla Model} & \multicolumn{1}{l|}{Source} \\ \hline
\multicolumn{1}{|l|}{Adaptive} & \multicolumn{1}{l|}{CIFAR-10} & \multicolumn{1}{l|}{74.1} & \multicolumn{1}{c|}{-} & \multicolumn{1}{c|}{$w_{c}=1, w_{q}=0, w_{\epsilon}=1$} & \multicolumn{1}{c|}{U} & \multicolumn{1}{l|}{ResNet56} & \multicolumn{1}{c|}{~\cite{mahmood2020beware}} \\ \hline
\multicolumn{1}{|l|}{Adaptive} & \multicolumn{1}{l|}{CIFAR-10} & \multicolumn{1}{l|}{22.3} & \multicolumn{1}{c|}{-} & \multicolumn{1}{c|}{$w_{c}=1, w_{q}=0, w_{\epsilon}=1$} & \multicolumn{1}{c|}{T} & \multicolumn{1}{l|}{ResNet56} & \multicolumn{1}{c|}{~\cite{mahmood2020beware}} \\ \hline
\multicolumn{1}{|l|}{DaST} & \multicolumn{1}{l|}{MNIST} & \multicolumn{1}{l|}{29.18} & \multicolumn{1}{c|}{-} & \multicolumn{1}{c|}{$w_{c}=1, w_{q}=0, w_{\epsilon}=1$} & \multicolumn{1}{c|}{U} & \multicolumn{1}{l|}{CNN} & \multicolumn{1}{c|}{~\cite{dast}} \\ \hline
\multicolumn{1}{|l|}{DaST} & \multicolumn{1}{l|}{MNIST} & \multicolumn{1}{l|}{57.22} & \multicolumn{1}{c|}{-} & \multicolumn{1}{c|}{$w_{c}=1, w_{q}=0, w_{\epsilon}=1$} & \multicolumn{1}{c|}{U} & \multicolumn{1}{l|}{CNN} & \multicolumn{1}{c|}{~\cite{dast}} \\ \hline
\multicolumn{1}{|l|}{DaST} & \multicolumn{1}{l|}{MNIST} & \multicolumn{1}{l|}{64.61} & \multicolumn{1}{c|}{-} & \multicolumn{1}{c|}{$w_{c}=1, w_{q}=0, w_{\epsilon}=1$} & \multicolumn{1}{c|}{U} & \multicolumn{1}{l|}{CNN} & \multicolumn{1}{c|}{~\cite{dast}} \\ \hline
\multicolumn{1}{|l|}{DaST} & \multicolumn{1}{l|}{MNIST} & \multicolumn{1}{l|}{96.36} & \multicolumn{1}{c|}{-} & \multicolumn{1}{c|}{$w_{c}=1, w_{q}=0, w_{\epsilon}=1$} & \multicolumn{1}{c|}{U} & \multicolumn{1}{l|}{CNN} & \multicolumn{1}{c|}{~\cite{dast}} \\ \hline
\multicolumn{1}{|l|}{DaST} & \multicolumn{1}{l|}{CIFAR-10} & \multicolumn{1}{l|}{19.78} & \multicolumn{1}{c|}{-} & \multicolumn{1}{c|}{$w_{c}=1, w_{q}=0, w_{\epsilon}=1$} & \multicolumn{1}{c|}{T} & \multicolumn{1}{l|}{VGG-16} & \multicolumn{1}{c|}{~\cite{dast}} \\ \hline
\multicolumn{1}{|l|}{DaST} & \multicolumn{1}{l|}{CIFAR-10} & \multicolumn{1}{l|}{20.22} & \multicolumn{1}{c|}{-} & \multicolumn{1}{c|}{$w_{c}=1, w_{q}=0, w_{\epsilon}=1$} & \multicolumn{1}{c|}{T} & \multicolumn{1}{l|}{VGG-16} & \multicolumn{1}{c|}{~\cite{dast}} \\ \hline
\multicolumn{1}{|l|}{DaST} & \multicolumn{1}{l|}{CIFAR-10} & \multicolumn{1}{l|}{28.42} & \multicolumn{1}{c|}{-} & \multicolumn{1}{c|}{$w_{c}=1, w_{q}=0, w_{\epsilon}=1$} & \multicolumn{1}{c|}{U} & \multicolumn{1}{l|}{VGG-16} & \multicolumn{1}{c|}{~\cite{dast}} \\ \hline
\multicolumn{1}{|l|}{DaST} & \multicolumn{1}{l|}{CIFAR-10} & \multicolumn{1}{l|}{59.71} & \multicolumn{1}{c|}{-} & \multicolumn{1}{c|}{$w_{c}=1, w_{q}=0, w_{\epsilon}=1$} & \multicolumn{1}{c|}{U} & \multicolumn{1}{l|}{VGG-16} & \multicolumn{1}{c|}{~\cite{dast}} \\ \hline
\multicolumn{1}{|l|}{Po+TI (Trip)} & \multicolumn{1}{l|}{ImageNet} & \multicolumn{1}{l|}{39.5} & \multicolumn{1}{c|}{-} & \multicolumn{1}{c|}{$w_{c}=1, w_{q}=0, w_{\epsilon}=1$} & \multicolumn{1}{c|}{T} & \multicolumn{1}{l|}{Inception v3} & \multicolumn{1}{c|}{~\cite{towards_transfer}} \\ \hline
\multicolumn{1}{|l|}{Po+TI (Trip)} & \multicolumn{1}{l|}{ImageNet} & \multicolumn{1}{l|}{39.3} & \multicolumn{1}{c|}{-} & \multicolumn{1}{c|}{$w_{c}=1, w_{q}=0, w_{\epsilon}=1$} & \multicolumn{1}{c|}{T} & \multicolumn{1}{l|}{ResNet50} & \multicolumn{1}{c|}{~\cite{towards_transfer}} \\ \hline
\multicolumn{8}{c}{Non-traditional   Attacks} \\ \hline
\multicolumn{1}{|l|}{Attack Name} & \multicolumn{1}{l|}{Dataset} & \multicolumn{1}{l|}{ASR} & \multicolumn{1}{l|}{Avg Queries} & \multicolumn{1}{c|}{Adv Threat Model} & \multicolumn{1}{l|}{Target} & \multicolumn{1}{l|}{Vanilla Model} & \multicolumn{1}{l|}{Source} \\ \hline
\multicolumn{1}{|l|}{CornerSearch} & \multicolumn{1}{l|}{MNIST} & \multicolumn{1}{c|}{97.38} & \multicolumn{1}{c|}{-} & \multicolumn{1}{c|}{$w_{c}=1, w_{q}=0, w_{\epsilon}=0$} & \multicolumn{1}{c|}{U} & \multicolumn{1}{l|}{NiN} & \multicolumn{1}{c|}{~\cite{croce2019sparse}} \\ \hline
\multicolumn{1}{|l|}{CornerSearch} & \multicolumn{1}{l|}{CIFAR-10} & \multicolumn{1}{c|}{99.56} & \multicolumn{1}{c|}{-} & \multicolumn{1}{c|}{$w_{c}=1, w_{q}=0, w_{\epsilon}=0$} & \multicolumn{1}{c|}{U} & \multicolumn{1}{l|}{NiN} & \multicolumn{1}{c|}{~\cite{croce2019sparse}} \\ \hline
\multicolumn{1}{|l|}{ColorFool} & \multicolumn{1}{l|}{CIFAR-10} & \multicolumn{1}{c|}{41.5} & \multicolumn{1}{c|}{-} & \multicolumn{1}{c|}{$w_{c}=1, w_{q}=0, w_{\epsilon}=0$} & \multicolumn{1}{c|}{U} & \multicolumn{1}{l|}{ResNet50} & \multicolumn{1}{c|}{~\cite{colorfool}} \\ \hline
\multicolumn{1}{|l|}{ColorFool} & \multicolumn{1}{l|}{ImageNet} & \multicolumn{1}{c|}{22.3} & \multicolumn{1}{c|}{-} & \multicolumn{1}{c|}{$w_{c}=1, w_{q}=0, w_{\epsilon}=0$} & \multicolumn{1}{c|}{U} & \multicolumn{1}{l|}{ResNet50} & \multicolumn{1}{c|}{~\cite{colorfool}} \\ \hline
\multicolumn{1}{|l|}{Patch Attack N4 4\%} & \multicolumn{1}{l|}{ImageNet} & \multicolumn{1}{c|}{99.7} & \multicolumn{1}{c|}{1137} & \multicolumn{1}{c|}{$w_{c}=1, w_{q}=0, w_{\epsilon}=0$} & \multicolumn{1}{c|}{U} & \multicolumn{1}{l|}{ResNet50} & \multicolumn{1}{c|}{~\cite{yang2020patchattack}} \\ \hline
\multicolumn{1}{|l|}{Patch Attack N8 2\%} & \multicolumn{1}{l|}{ImageNet} & \multicolumn{1}{c|}{99.7} & \multicolumn{1}{c|}{983} & \multicolumn{1}{c|}{$w_{c}=1, w_{q}=0, w_{\epsilon}=0$} & \multicolumn{1}{c|}{U} & \multicolumn{1}{l|}{ResNet50} & \multicolumn{1}{c|}{~\cite{yang2020patchattack}} \\ \hline
\multicolumn{1}{|l|}{Patch Attack N10 4\%} & \multicolumn{1}{l|}{ImageNet} & \multicolumn{1}{c|}{99.7} & \multicolumn{1}{c|}{8643} & \multicolumn{1}{c|}{$w_{c}=1, w_{q}=0, w_{\epsilon}=0$} & \multicolumn{1}{c|}{T} & \multicolumn{1}{l|}{ResNet50} & \multicolumn{1}{c|}{~\cite{yang2020patchattack}} \\ \hline
\multicolumn{1}{|l|}{Patch Attack N10 10\%} & \multicolumn{1}{l|}{ImageNet} & \multicolumn{1}{c|}{100} & \multicolumn{1}{c|}{3747} & \multicolumn{1}{c|}{$w_{c}=1, w_{q}=0, w_{\epsilon}=0$} & \multicolumn{1}{c|}{T} & \multicolumn{1}{l|}{ResNet50} & \multicolumn{1}{c|}{~\cite{yang2020patchattack}} \\ \hline
\end{tabular}
\caption{Transfer based and non-traditional attacks on various datasets (MNIST, CIFAR-10 and ImageNet). NiN stands for Network-in-Network. \label{tbl:nt}}
\end{table*}
\section{Conclusion}
\label{sec:conclusion}

Adversarial machine learning is advancing at a fast pace, with new attack papers being proposed every year. In light of these recent developments, we have surveyed the current state-of-the-art black-box attack and have provided three major contributions. First, our survey covers 20 new attack papers with detailed summaries, mathematics and attack explanations. Our second contribution is a categorization of these attacks into four different types, score based, decision based, transfer based and non-traditional attacks. This organization assists new readers in comprehending the field and helps current researchers understand where each new attacks fits in the rapidly growing black-box adversarial machine learning literature. Lastly, we offer a new mathematical framework for defining the adversarial threat model. Our new framework provides a convenient and efficient way to quickly determine when attack success rates from different attacks can be compared. Without this framework, we have shown that directly comparing attack success rates from different papers with different threat models can lead to highly misleading conclusions. Overall, our work and comparative evaluations provide insight, organization and systemization to the developing field of adversarial machine learning. 
\bibliographystyle{ACM-Reference-Format}
\bibliography{ref}
\end{document}